\documentclass{nature}

\newcounter{firstbib}
\newcounter{secondbib}

\usepackage{graphicx}
\usepackage{epsfig}
\usepackage[font={small}]{caption}
\usepackage{hyperref}
\usepackage{amsmath}
\usepackage{amssymb}
\usepackage{lineno}
\usepackage{tikz}
\tikzset{every picture/.style={/utils/exec={\sffamily}}}

\usepackage[symbol]{footmisc}

\usepackage{xcolor}

\def\Msun{{\rm M}_\odot}
\def\Lsun{{\rm L}_\odot}

\def\gtorder{\mathrel{\raise.3ex\hbox{$>$}\mkern-14mu
             \lower0.6ex\hbox{$\sim$}}}
\def\ltorder{\mathrel{\raise.3ex\hbox{$<$}\mkern-14mu
             \lower0.6ex\hbox{$\sim$}}}

\def\Angstrom{\textup{\AA}}

\def\farcs{\hbox{$.\!\!^{\prime\prime}$}}
\def\farcm{\hbox{$.\!\!^{\prime}$}}

\newcommand{\aap}{Astron. Astrophys.}
\newcommand{\araa}{Ann. Rev. Astron. Astrophys.}
\newcommand{\apj}{Astrophys. J.}
\newcommand{\aj}{Astron. J.}
\newcommand{\apjl}{Astrophys. J. Lett.}
\newcommand{\apjs}{Astrophys. J. Suppl.}
\newcommand{\nat}{Nature}
\newcommand{\natp}{Nat. Phys.}
\newcommand{\pasp}{Publ. Astron. Soc. Pac.}
\newcommand{\pasa}{Publ. Astron. Soc. Aust.}
\newcommand{\pasj}{Publ. Astron. Soc. Jpn.}
\newcommand{\mnras}{Mon. Not. R. Astron. Soc.}

\newcommand{\apss}{Astrophys. Space Sci.}  

\title{The electron-capture origin of supernova 2018zd}

\author{
Daichi Hiramatsu$^{\ref{af:LCO},\ref{af:UCSB}*}$,
D. Andrew Howell$^{\ref{af:LCO},\ref{af:UCSB}}$,
Schuyler D. Van Dyk$^{\ref{af:Spitzer}}$,
Jared A. Goldberg$^{\ref{af:UCSB}}$,
Keiichi Maeda$^{\ref{af:Kyoto},\ref{af:IPMU}}$,
Takashi J. Moriya$^{\ref{af:NAOJ}, \ref{af:Monash}}$,
Nozomu Tominaga$^{\ref{af:Konan}, \ref{af:IPMU}, \ref{af:NAOJ}}$,
Ken’ichi Nomoto$^{\ref{af:IPMU}}$,
Griffin Hosseinzadeh$^{\ref{af:CfA}}$,
Iair Arcavi$^{\ref{af:TelAviv}, \ref{af:CIFAR}}$,
Curtis McCully$^{\ref{af:LCO}, \ref{af:UCSB}}$,
Jamison Burke$^{\ref{af:LCO},\ref{af:UCSB}}$,
K. Azalee Bostroem$^{\ref{af:UCD}}$,
Stefano Valenti$^{\ref{af:UCD}}$,
Yize Dong$^{\ref{af:UCD}}$,
Peter J. Brown$^{\ref{af:AM}}$,
Jennifer E. Andrews$^{\ref{af:Arizona}}$,
Christopher Bilinski$^{\ref{af:Arizona}}$,
G. Grant Williams$^{\ref{af:Arizona}, \ref{af:MMT}}$,
Paul S. Smith$^{\ref{af:Arizona}}$,
Nathan Smith$^{\ref{af:Arizona}}$,
David J. Sand$^{\ref{af:Arizona}}$,
Gagandeep S. Anand$^{\ref{af:IPAC}, \ref{af:IfA}}$,
Chengyuan Xu$^{\ref{af:UCSBMedia}}$,
Alexei V. Filippenko$^{\ref{af:Berkeley},\ref{af:Miller}}$,
Melina C. Bersten$^{\ref{af:IALP},\ref{af:LaPlata},\ref{af:IPMU}}$,
Gast\'{o}n Folatelli$^{\ref{af:IALP},\ref{af:LaPlata},\ref{af:IPMU}}$,
Patrick L. Kelly$^{\ref{af:Minnesota}}$,
Toshihide Noguchi$^{\ref{af:Noguchi}}$,
\&
Koichi Itagaki$^{\ref{af:Itagaki}}$
\\
\normalsize{*Corresponding author. Email: dhiramatsu@lco.global}
}

\begin{document}

\maketitle

\begin{affiliations}
 \item Las Cumbres Observatory, 6740 Cortona Drive, Suite 102, Goleta, CA 93117-5575, USA\label{af:LCO}
 \item Department of Physics, University of California, Santa Barbara, CA 93106-9530, USA\label{af:UCSB}
 \item Caltech/Spitzer Science Center, Caltech/IPAC, Mailcode 100-22, Pasadena, CA 91125, USA\label{af:Spitzer}
 \item Department of Astronomy, Kyoto University, Kitashirakawa-Oiwake-cho, Sakyo-ku, Kyoto 606-8502, Japan\label{af:Kyoto}
 \item Kavli Institute for the Physics and Mathematics of the Universe (WPI), The University of Tokyo Institutes for Advanced Study, The University of Tokyo, 5-1-5
Kashiwanoha, Kashiwa, Chiba 277-8583, Japan\label{af:IPMU}
 \item National Astronomical Observatory of Japan, National Institutes of Natural Sciences, 2-21-1 Osawa, Mitaka, Tokyo 181-8588, Japan\label{af:NAOJ}
 \item School of Physics and Astronomy, Faculty of Science, Monash University, Clayton, Victoria 3800, Australia\label{af:Monash}
 \item Department of Physics, Faculty of Science and Engineering, Konan University, 8-9-1 Okamoto, Kobe, Hyogo 658-8501, Japan\label{af:Konan}
 \item Center for Astrophysics \textbar{} Harvard \& Smithsonian, 60 Garden Street, Cambridge, MA 02138-1516, USA\label{af:CfA}
 \item The School of Physics and Astronomy, Tel Aviv University, Tel Aviv 69978, Israel\label{af:TelAviv}
 \item CIFAR Azrieli Global Scholars program, CIFAR, Toronto, Canada\label{af:CIFAR}
 \item Department of Physics, University of California, 1 Shields Ave, Davis, CA 95616-5270, USA\label{af:UCD}
 \item Mitchell Institute for Fundamental Physics and Astronomy, Texas A\&M University, College Station, TX 77843, USA\label{af:AM}
 \item Steward Observatory, University of Arizona, 933 North Cherry Avenue, Tucson, AZ 85721-0065, USA\label{af:Arizona}
 \item MMT Observatory, P.O. Box 210065, Tucson, AZ 85721-0065, USA\label{af:MMT}
 \item Infrared Processing and Analysis Center, California Institute of Technology, Pasadena, CA 91125, USA\label{af:IPAC}
 \item Institute for Astronomy, University of Hawai`i, 2680 Woodlawn Drive, Honolulu, HI 96822, USA\label{af:IfA}
 \item Media Arts and Technology, University of California, Santa Barbara, CA 93106-6065, USA\label{af:UCSBMedia}
 \item Department of Astronomy, University of California, Berkeley, CA 94720-3411, USA\label{af:Berkeley}
 \item Miller Institute for Basic Research in Science, University of California, Berkeley, CA 94720, USA\label{af:Miller}
 \item Instituto de Astrof\'{i}sica de La Plata (IALP), CONICET, Argentina\label{af:IALP}
 \item Facultad de Ciencias Astron\'{o}micas y Geof\'{i}sicas, Universidad Nacional de La Plata, Paseo del Bosque, B1900FWA, La Plata, Argentina\label{af:LaPlata}
 \item School of Physics and Astronomy, University of Minnesota, 116 Church Street SE, Minneapolis, MN 55455, USA\label{af:Minnesota}
 \item Noguchi Astronomical Observatory, Katori, Chiba 287-0011, Japan\label{af:Noguchi}
 \item Itagaki Astronomical Observatory, Yamagata, Yamagata 990-2492, Japan\label{af:Itagaki}
\end{affiliations}




\clearpage

\begin{abstract}

In the transitional mass range ($\sim$ 8--10 solar masses) between white dwarf formation and iron core-collapse supernovae, stars are expected to produce an electron-capture supernova. 
Theoretically, these progenitors are thought to be super-asymptotic giant branch stars with a degenerate O+Ne+Mg core, and electron capture onto Ne and Mg nuclei should initiate core collapse\cite{Miyaji1980,Nomoto1982,Nomoto1984,Nomoto1987}.
However, no supernovae have unequivocally been identified from an electron-capture origin, partly because of uncertainty in theoretical predictions.  
Here we present six indicators of electron-capture supernovae and show that supernova 2018zd is the only known supernova having strong evidence for or consistent with all six: progenitor identification, circumstellar material, chemical composition\cite{Poelarends2008,Jones2013,Doherty2017}, explosion energy, light curve, and nucleosynthesis\cite{Kitaura2006,Janka2008,Tominaga2013,Wanajo2009,Jerkstrand2018}. 
For supernova 2018zd, we infer a super-asymptotic giant branch progenitor based on the faint candidate in the pre-explosion images and the chemically-enriched circumstellar material revealed by the early ultraviolet colours and flash spectroscopy. The light-curve morphology and nebular emission lines can be explained by the low explosion energy and neutron-rich nucleosynthesis produced in an electron-capture supernova.
This identification provides insights into the complex stellar evolution, supernova physics, cosmic nucleosynthesis, and remnant populations in the transitional mass range.

\end{abstract}


On 2018 Mar. 2.49 (UT dates are used throughout), we discovered AT~2018zd\cite{Itagaki2018} at an unfiltered optical magnitude of $17.8$ in the outskirts of NGC 2146 (redshift $z=0.002979$; ref.~\cite{de1991}), where pre-explosion \textit{Hubble Space Telescope} (\textit{HST}) and \textit{Spitzer Space Telescope} images yield a faint progenitor candidate (Extended Data Figs.~\ref{EDfig:discovery} and \ref{EDfig:progenitor_sed}, and Methods). 
Combined with our pre-discovery detection at $18.1$ mag on 2018 Mar. 1.54, we estimate an explosion epoch of 2018 Mar. $1.4 \pm 0.1$ ($\sim3$ hr before the first detection; Extended Data Fig.~\ref{EDfig:allLC}) and use it as a reference epoch for all phases. 
At 4.9 days post explosion, we classified AT~2018zd as a young Type~II (hydrogen-rich) supernova (SN), designating it SN~2018zd\cite{Arcavi2018}. Over time, SN~2018zd developed a plateau and broad Balmer-series P Cygni lines in the optical light curves and spectra (respectively), further classifying it as a Type~II-P (plateau) SN (Extended Data Figs.~\ref{EDfig:allLC} and \ref{EDfig:allspec}). 
The luminosity distance of NGC 2146 is uncertain, ranging from $11$\,Mpc to $18$\,Mpc in the literature\cite{Adamo2012}. Thus, we apply the standard candle method and adopt a distance of $9.6\pm1.0$\,Mpc (Methods). Because of the wide distance range, we focus mainly on distance-independent measurements. 

Unlike Fe core-collapse (CC) SN explosions of red supergiant (RSG) stars, electron-capture (EC) SN explosions of super-asymptotic giant branch (SAGB) stars are robustly realised by first-principle simulations, facilitated by the steep density gradient outside the degenerate core. Simulations consistently predict explosion energy ($\sim2\times10^{50}$\,erg) and $^{56}$Ni yield ($\sim3\times10^{-3}\,M_\odot$, with an upper limit $\lesssim10^{-2}\,M_\odot$) that are an order of magnitude lower than those observed for typical Fe CCSNe\cite{Kitaura2006, Janka2008, Wanajo2009}, but are consistent within the lowest-mass Fe CCSNe (Supplementary Information). 
Despite the low explosion energy, the low mass and large radius of an SAGB star result in a light-curve morphology virtually identical to that of Type~II-P SNe, except for a larger drop ($\sim4$\,mag) from the plateau to the radioactive decay tail, owing to the low $^{56}$Ni production\cite{Tominaga2013}.

\begin{figure}
\centering
\includegraphics[width=0.99\textwidth]{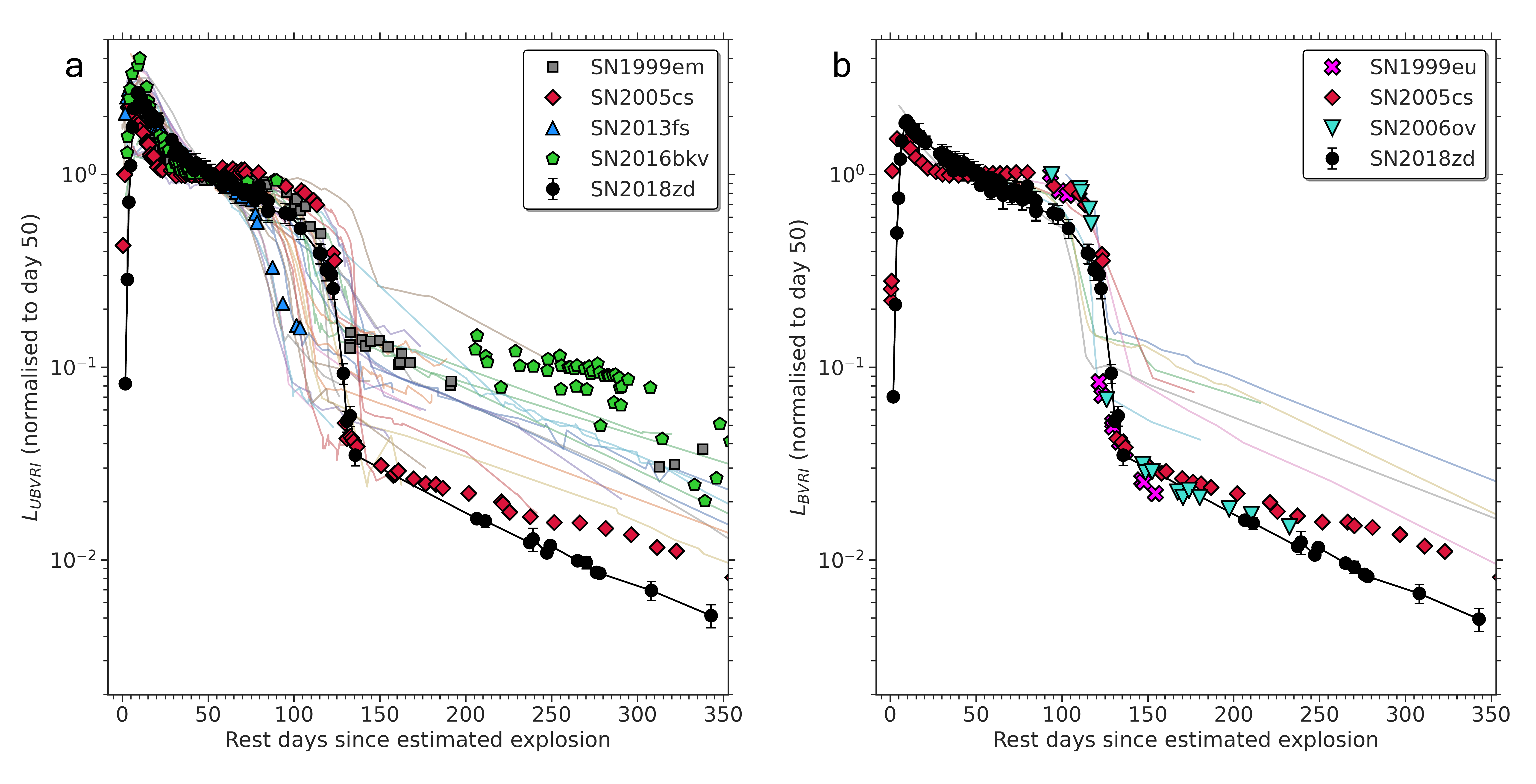}
\caption{
\textbf{Normalised pseudobolometric light curves.}
\textbf{a,} Comparison of the normalised pseudobolometric ({\it UBVRI}; Supplementary Information) light curve of SN~2018zd with a  well-observed Type~II SN sample\cite{Valenti2016} (transparent lines), including archetypal SN~1999em, low-luminosity SN~2005cs, and early-flash SN~2013fs, along with low-luminosity and early-flash SN~2016bkv\cite{Hosseinzadeh2018}. 
\textbf{b,} Comparison of the normalised pseudobolometric ({\it BVRI}) light curve of SN~2018zd with a low-luminosity Type~II-P SN sample\cite{Spiro2014}, including SNe~1999eu and 2006ov with the largest plateau drops ever (to our knowledge). 
Error bars denote $1\sigma$ uncertainties. 
Because of the distance uncertainty of SN~2018zd, we normalise each light curve to day 50 and make the comparisons distance independent. 
SN~2018zd shows the largest plateau drop and is comparable to that of SNe~1999eu and 2006ov, indicating an intrinsically low $^{56}$Ni production.
\label{fig:L50}
}
\end{figure}

Among a sample of well-observed Type~II SN light curves\cite{Valenti2016} (Fig.~\ref{fig:L50}), SN~2018zd fits in the Type~II-P morphology and displays the largest plateau drop ($\sim 3.8$ mag). 
Even among a sample of low-luminosity Type~II-P SNe\cite{Spiro2014} that often show larger plateau drops than other Type~II subclasses (Fig.~\ref{fig:L50}), SN~2018zd is comparable to SNe 1999eu and 2006ov with the largest drops ever observed, indicating an intrinsically low $^{56}$Ni production. 
For SNe 1999eu and 2006ov, the lack of additional data prevents the investigations of other ECSN indicators; the light curves alone cannot be conclusive evidence (see Methods for the light-curve degeneracy).
The tail decline rate of SN~2018zd is consistent with the $^{56}$Co heating rate, and an estimated $^{56}$Ni mass is $(8.6\pm 0.5)\times10^{-3}\,M_\odot$ at the assumed luminosity distance of $9.6$\,Mpc (Extended Data Fig.~\ref{EDfig:allLC}). This is larger than the canonical $^{56}$Ni yield for ECSNe, but still within the upper limit (see also Supplementary Information for the effect of distance uncertainty).

As SAGB stars are thought to have mass-loss rates ($\dot{M} \approx 10^{-4}\,M_\odot$\,yr$^{-1}$) a few orders of magnitude higher than those of RSG stars of similar initial mass\cite{Poelarends2008}, the circumstellar material (CSM) density is expected to be a few orders of magnitude higher, as it scales as $\rho_{\text{CSM}} \propto \dot{M}/v_{\rm wind}$, assuming constant-wind mass loss with similar SAGB and RSG wind velocities\cite{Moriya2014} $v_{\rm wind}$. 
Compared with RSG stars, the CSM composition of SAGB stars can be He-, C-, and N-rich, but O-poor, depending on the efficiency of the SAGB dredge-up and dredge-out that bring the partial H- and He-burning products to the stellar surface\cite{Jones2013, Doherty2017}.

\begin{figure}
\centering
\includegraphics[width=0.99\textwidth]{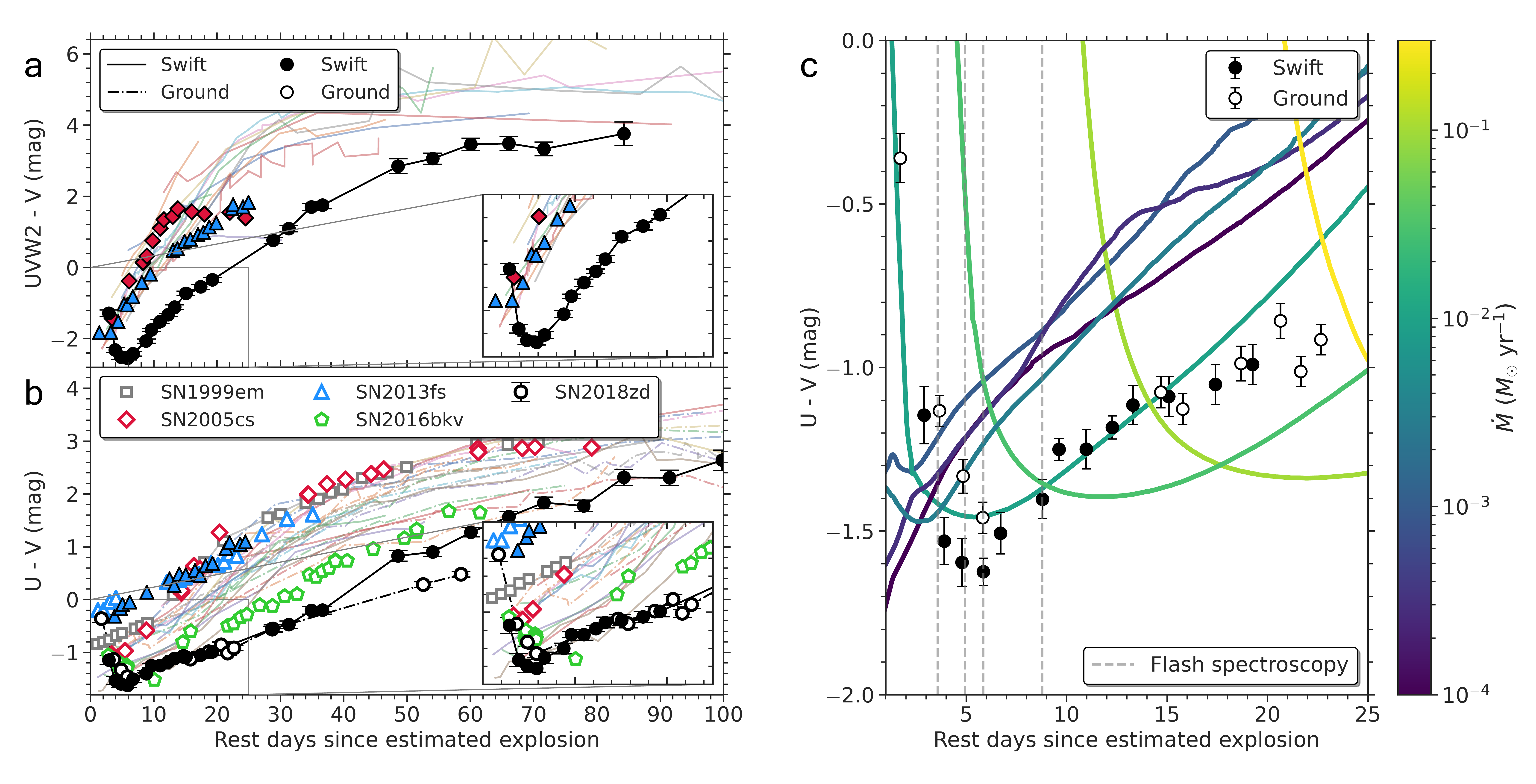}
\caption{
\textbf{UV colour light curves.}
\textbf{a, b,} Comparison of the UV colours of SN~2018zd with the sample as in Fig.~\ref{fig:L50}a. The panels show \textit{Swift} $UVW2-V$ (\textbf{a}) and \textit{Swift} and ground-based $U-V$ colour evolution (\textbf{b}).
Note the pronounced sharp blueward colour evolution of SN~2018zd over the first $\sim 5$\,d, shown in the insets, suggesting a possible delayed shock-breakout through dense CSM. 
\textbf{c,} Comparison of the $U-V$ colour of SN~2018zd with our \texttt{MESA}+\texttt{STELLA} CSM models (Methods and Extended Data Fig.~\ref{EDfig:LCscaling}) assuming a typical constant wind velocity of $20$\,km\,s$^{-1}$ (Fig.~\ref{fig:flash}), colour-coded by the mass-loss rate. 
To reproduce the observed early blueward evolution, $\dot{M} \approx 0.01\,M_\odot$\,yr$^{-1}$ for the last $\sim 10$\,yr before the explosion is required. The observed flash-spectroscopy epochs (Fig.~\ref{fig:flash}) are marked by the vertical dashed lines and are consistent with the blueward colour evolution.
Error bars denote $1\sigma$ uncertainties. 
\label{fig:uvUV}
}
\end{figure}

In a sample of Type~II SN ultraviolet (UV) colours\cite{Valenti2016}, SN~2018zd stands out, reaching the minimum in $U-V$ colour (that is, becoming bluer until) $\sim5$\,d after the explosion (Fig.~\ref{fig:uvUV}), which suggests a possible delayed shock-breakout through dense CSM.
In such a case, a photosphere initially forms inside the unshocked optically-thick CSM\cite{Moriya2018}; this provides an additional power source leading to the bluer colour when the shock front is propagating through the CSM (see Extended Data Fig.~\ref{EDfig:allvel} for the same effect on the photospheric velocity). 
Our \texttt{MESA}+\texttt{STELLA} CSM light-curve models (Methods and Extended Data Fig.~\ref{EDfig:LCscaling}) show that $\dot{M} \approx 0.01\,M_\odot$\,yr$^{-1}$ for the last $\sim10$\,yr before the explosion is required to reproduce the early-time $U-V$ colour of SN~2018zd, assuming a typical constant\cite{Moriya2014} $v_{\text{wind}} = 20$\,km\,s$^{-1}$ (Fig.~\ref{fig:uvUV}). Since the estimated mass loss is a few orders of magnitude greater than that expected from SAGB or RSG winds, it is probably dominated by eruptive events\cite{Poelarends2008, Jones2013}. 

\begin{figure}
\centering
\includegraphics[width=0.99\textwidth]{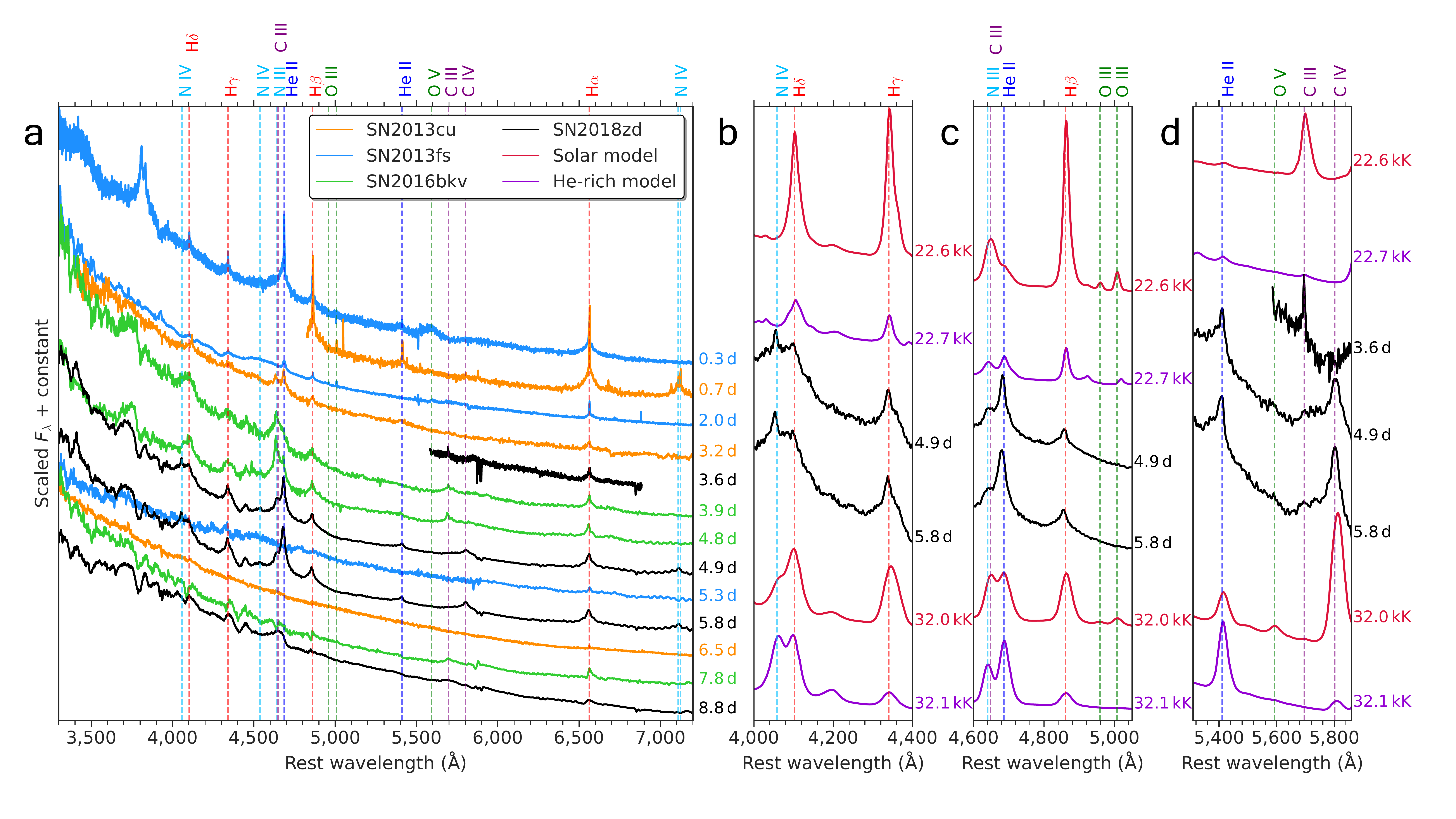}
\caption{
\textbf{Flash spectral time series.}
\textbf{a,} Comparison of the flash spectral time series of SN~2018zd with that of well-sampled Type~II-P SNe 2013fs\cite{Yaron2017} and 2016bkv\cite{Hosseinzadeh2018}, and Type~IIb (mostly stripped H-rich envelope) SN~2013cu\cite{Gal-Yam2014}. 
SN~2018zd exhibits the persistent flash features ($\gtrsim9$ d), while most of the flash features in the other SNe disappear within $\sim5$\,d after the explosion. 
\textbf{b--d,} Comparison of the flash spectral time series at three different zoomed-in wavelength regions of SN~2018zd with the scaled and resampled flash spectral models of solar abundance ($X_{\rm H}=0.70$, $X_{\rm He}=0.28$, $X_{\rm C}=3.02\times10^{-3}$, $X_{\rm N}=1.18\times10^{-3}$, $X_{\rm O}=9.63\times10^{-3}$) and He-rich ($X_{\rm H}=0.18$, $X_{\rm He}=0.80$, $X_{\rm C}=5.58\times10^{-5}$, $X_{\rm N}=8.17\times10^{-3}$, $X_{\rm O}=1.312\times10^{-4}$) atmosphere with $\dot{M}_{\rm{mod}}=3\times10^{-3}\,M_\odot$\,yr$^{-1}$ and $v_{\rm{mod}}=150$\,km\,s$^{-1}$ (the densest CSM with the finest temperature grid spacing)\cite{Boian2019}. 
The temperatures are constrained to be within $\sim20$,$000$\,K (at 3.6\,d) to $30$,$000$\,K (at 4.9--5.8\,d) from the \texttt{MESA}+\texttt{STELLA} UV-colour models (Fig.~\ref{fig:uvUV}). 
The observed features are expected to be narrower and stronger if resolved, as $\rho_{\text{obs}}/\rho_{\text{mod}}=(\dot{M}_{\text{obs}}/v_{\text{obs}})/(\dot{M}_{\text{mod}}/v_{\text{mod}})=25$ with $\dot{M}_{\text{obs}}=0.01\,M_\odot$\,yr$^{-1}$ from the UV colours and assuming $v_{\text{obs}}=20$\,km\,s$^{-1}$ (the wind P Cygni components of SN~2018zd are not resolved, only giving an upper-limit $v_{\rm obs} < 76.3$\,km\,s$^{-1}$ from the highest spectral resolution of C~{\sc iii} $\lambda$5696 at 3.6\,d). 
On the basis of the model comparisons, the line ratios of N~{\sc iv}/H$\delta$ $>1$ (\textbf{b}) and He~{\sc ii}/H$\beta$ $>1$ (\textbf{c}), the transition from C~{\sc iii} to C~{\sc iv} (\textbf{d}), and the lack of O~{\sc iii} and O~{\sc v} lines (\textbf{c, d}) observed in SN~2018zd suggest He-, C-, and N-rich, but O-poor CSM composition.
\label{fig:flash}
}
\end{figure}

Consistent with the possible delayed shock breakout seen in the early UV colour, SN~2018zd exhibits unusually persistent ($\gtrsim9$ d) flash features, reaching the highest ionisation states at $\sim5$\,d after the explosion (Fig.~\ref{fig:flash}). 
The strengths of flash features depend on the photospheric temperature, CSM density, and CSM abundance\cite{Yaron2017,Boian2019,Boian2020}. 
We constrain the photospheric temperature and CSM density of SN~2018zd by the \texttt{MESA}+\texttt{STELLA} UV-colour models. Then we use emission-line intensity ratios as diagnostics of CSM abundance by comparing with the flash spectral models of solar-abundance and He-rich atmospheres\cite{Boian2019} (Fig.~\ref{fig:flash}; note that the line ratios are not well reproduced by either solar-abundance or He-rich models alone\cite{Boian2020}, and a mixture of both with higher density needs to be modelled for a more detailed abundance analysis). On the basis of the model comparisons, we estimate He- C-, and N-rich, but O-poor CSM mass fractions of $X_{\rm He}\approx0.3$--$0.8$, $X_{\rm C}\approx3\times10^{-3}$, $X_{\rm N}\approx8\times10^{-3}$, and $X_{\rm O}\approx10^{-4}$, which is more consistent with an SAGB than an RSG atmosphere\cite{Jones2013, Doherty2017}.  

Since the core composition and explosion nucleosynthesis are different from Fe CCSNe (but see Supplementary Information for some caveats on the low-mass end), ECSNe are expected to show distinct nebular spectral features: stronger Ni than Fe lines due to a more stable $^{58}$Ni yield than radioactive $^{56}$Ni (a parent nuclide of $^{56}$Fe) from the innermost neutron-rich ejecta (electron fraction $Y_e\lesssim0.49$)\cite{Wanajo2009,Jerkstrand2018}; weak O, Mg, and Fe lines owing to the thin ($\sim0.01\,M_\odot$) O+C shell, which is further burned to Fe-group elements\cite{Kitaura2006,Janka2008}; and weak C lines because of the efficient dredge-up/out reducing most of the He-rich layer before the explosion\cite{Poelarends2008,Jones2013,Doherty2017} (N lines are hard to be constrained in Type~II-P SNe, as [N~{\sc ii}] $\lambda\lambda$6548, 6583 are hidden by strong H$\alpha$).
For low-mass progenitors ($\lesssim12\,M_\odot$), a low line ratio of [O~{\sc i}]/[Ca~{\sc ii}] is expected owing to the low O-core mass\cite{Jerkstrand2012,Maeda2007}.

\begin{figure}
\centering
\includegraphics[width=0.99\textwidth]{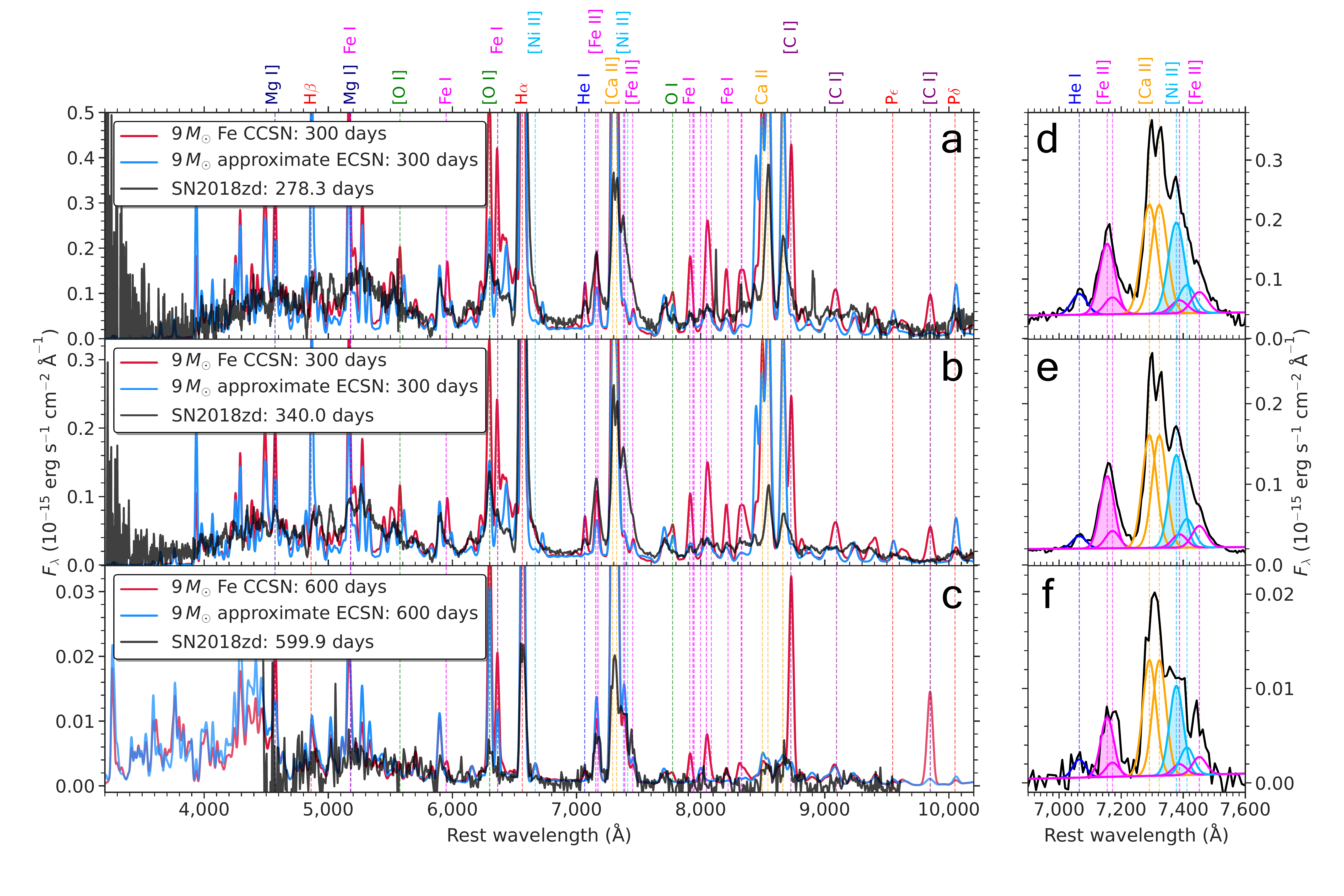}
\caption{
\textbf{Nebular spectral time series.}
\textbf{a--c,} Comparison of the nebular spectral time series at three different epochs of SN~2018zd with the scaled (by $^{56}$Ni mass and phase) and resampled $9\,M_\odot$ Fe CCSN and `approximate ECSN' (excluding the He-core composition from the Fe CCSN) models\cite{Jerkstrand2018}. 
The large number of narrow lines at $\lesssim 5500$\,\AA\ and strong Ca lines in the models are known issues. The weak [Ni~{\sc ii}] lines in the models are from the primordial Ni (solar abundance), as Ni nucleosynthesis is not taken into account.
In ascending order of wavelength, note the weak Mg~{\sc i}] $\lambda$4571, Mg~{\sc i}] $\lambda$5180 + Fe~{\sc i} $\lambda$5180, [O~{\sc i}] $\lambda$5577, Fe~{\sc i} $\lambda$5950, [O~{\sc i}] $\lambda\lambda$6300, 6364 + Fe~{\sc i} $\lambda$6364, O~{\sc i} $\lambda$7774, Fe~{\sc i} cluster 7,900--8,500\,\AA, [C~{\sc i}] $\lambda$8727, [C~{\sc i}] $\lambda$9100, and [C~{\sc i}] $\lambda$9850, as well as the low line ratio of [O~{\sc i}] $\lambda\lambda$6300, 6364/[Ca~{\sc ii}] $\lambda\lambda$7291, 7323 $<1$ observed in SN~2018zd.
He~{\sc i} $\lambda$7065 is weaker in the approximate ECSN model than in the observed spectra; the emission from the dredged-up/out elements (for example, He and N) in the H-rich envelope may be underestimated by the model.
\textbf{d--f,} Simultaneous Gaussian fits to He~{\sc i} $\lambda$7065, [Fe~{\sc ii}] $\lambda$7155, [Fe~{\sc ii}] $\lambda$7172, [Ca~{\sc ii}] $\lambda$7291, [Ca~{\sc ii}] $\lambda$7323, [Ni~{\sc ii}] $\lambda$7378, [Fe~{\sc ii}] $\lambda$7388, [Ni~{\sc ii}] $\lambda$7412, and [Fe~{\sc ii}] $\lambda$7452 (Supplementary Information) at three different epochs.
Note the stronger [Ni~{\sc ii}] $\lambda$7378 (the blue shaded region) than [Fe~{\sc ii}] $\lambda$7155 (the pink shaded region), yielding [Ni~{\sc ii}]/[Fe~{\sc ii}] $= 1.3$--$1.6$.
The weak C, O, Mg, and Fe lines combined with the strong Ni lines observed in SN~2018zd are consistent with the ECSN chemical composition and nucleosynthesis.
\label{fig:nebular}
}
\end{figure}

True nebular spectral models of ECSNe are difficult to produce, but they can be approximated by removing the He core from an Fe CCSN simulation. Here we use such a model\cite{Jerkstrand2018}, which we call the `approximate ECSN' model. Comparison of the nebular spectra of SN~2018zd with the $9\,M_\odot$ models\cite{Jerkstrand2018} favours the approximate ECSN model over the Fe CCSN model, especially through the weak C, O, Mg, and Fe lines (Fig.~\ref{fig:nebular}). In addition, the low line ratio of [O~{\sc i}] $\lambda\lambda$6300, 6364/[Ca~{\sc ii}] $\lambda\lambda$7291, 7323 $<1$ observed in SN~2018zd indicates a low-mass progenitor. 

Although quantitative analysis to derive the masses of Ni and Fe requires detailed radiative-transfer simulations, we can obtain a rough estimate of the line ratio expected from ECSNe. 
For normal Fe CCSNe where Ni and Fe are dominantly produced in the same layer,  [Fe~{\sc ii}] overwhelms [Ni~{\sc ii}] in the emission from the innermost region\cite{Jerkstrand2012}. In ECSN models\cite{Wanajo2009}, however, there is a layer of Ni-rich (neutron-rich) material, emitting predominantly [Ni~{\sc ii}], inside the outer mixture of Ni and Fe. In this situation where the Ni-rich and Fe-rich regions are physically separated, [Ni~{\sc ii}]/[Fe~{\sc ii}] roughly reflects the mass ratio of Ni and Fe in the entire ejecta\cite{Maeda2007}, which is 1.3--3.0 in the ECSN models\cite{Wanajo2009}. 
The observed [Ni~{\sc ii}] $\lambda$7378/[Fe~{\sc ii}] $\lambda$7155 ratio of 1.3--1.6 in SN~2018zd (Fig.~\ref{fig:nebular}) is indeed within the expected range. 
In principle, clumping, fluorescence, and/or shock excitation could enhance [Ni~{\sc ii}] $\lambda$7378 such that [Ni~{\sc ii}] $\lambda$7378/[Fe~{\sc ii}] $\lambda$7155 overestimates the Ni/Fe mass ratio\cite{Hudgins1990}, but we leave a detailed theoretical study to future works. 

SN~2018zd fulfills the expected characteristics and is strong evidence for the existence of ECSNe and their progenitor SAGB stars (see Supplementary Information and Extended Data Fig.~\ref{EDfig:ECSNcan} for other previously suggested ECSN candidates). 
With SN~2018zd, we roughly estimate an ECSN rate of $0.6$--$8.5$\% of all CCSNe, corresponding to a narrow SAGB progenitor mass window of $\Delta M_{\rm SAGB} \approx 0.06$--$0.69$\,$M_\odot$ (Methods and Extended Data Fig.~\ref{EDfig:rate}).
Theoretically, the evolutionary path to SAGB stars is uncertain owing to the high sensitivity of nuclear burning on complex dredge-up/out and mass-loss mechanisms\cite{Doherty2017, Jones2013}, giving a variety of expected mass windows at different metallicities (for example, $\Delta M_{\rm SAGB}\approx0.2$--$1.4$\,$M_\odot$ at solar metallicity\cite{Poelarends2008}). 
Their final fate may vary from CC to thermonuclear ECSNe depending on the electron-capture rates and oxygen flame speed in the degenerate core\cite{Zha2019, Leung2020}, resulting in different nucleosynthetic yields and galactic chemical evolution\cite{Jones2019}.
The CC ECSNe are expected to leave low mass, spin, and kick-velocity neutron stars\cite{Gessner2018}, forming a low-mass peak ($\sim1.25\,M_\odot$) in neutron star mass distribution\cite{Schwab2010} and low-eccentricity ($\sim0.2$) gravitational wave source population\cite{Giacobbo2019}. 
Therefore, using SN~2018zd as an ECSN template, future statistical studies with homogeneous samples from large surveys will be able to further reveal the evolution of SAGB progenitors and the influence of ECSNe on the kinetic and chemical composition of the Universe.

%
%

%



 
\begin{addendum}

\item[Acknowledgements]
We are grateful to A.~Suzuki, T.~Takiwaki, T.~Nozawa, M.~Tanaka, C.~Kobayashi, R.~Ouchi, T.~Matsuoka, T.~Hayakawa, S.~I.~Blinnikov, K.~Chen, L.~Bildsten, and B.~Paxton for comments and discussions, to C.~P.~Guti{\'e}rrez and A.~Pastorello for sharing the velocity data of the Type~II SN sample and SN~2005cs (respectively), and to Peter~Il\'a\v{s} for creating the colour-composite image.

D.H., D.A.H., G.H., C.M., and J.B. were supported by the U.S. National Science Foundation (NSF) grants AST-1313484 and AST-1911225, as well as by the National Aeronautics and Space Administration (NASA) grant 80NSSC19kf1639. 
D.H. is thankful for support and hospitality by the Kavli Institute for the Physics and Mathematics of the Universe (IPMU) where many discussions of this work took place.
J.A.G. is supported by the NSF GRFP under grant 1650114.
K.M. acknowledges support by JSPS KAKENHI grants 20H00174, 20H04737, 18H04585, 18H05223, and 17H02864.
K.N.'s work and D. H.'s visit to Kavli IPMU have been supported by the World Premier International Research Center Initiative (WPI Initiative), MEXT, and JSPS KAKENHI grants JP17K05382 and JP20K04024, Japan.
I.A. is a CIFAR Azrieli Global Scholar in the Gravity and the Extreme Universe Program and acknowledges support from that program, from the Israel Science Foundation (grants 2108/18 and 2752/19), from the United States -- Israel Binational Science Foundation (BSF), and from the Israeli Council for Higher Education Alon Fellowship.
Research by K.A.B., S.V., and Y.D. is supported by NSF grant AST–1813176.
J.E.A. and N.S. receive support from NSF grant AST-1515559.
Research by D.J.S. is supported by NSF grants AST-1821967, 1821987, 1813708, 1813466, and 1908972.
G.S.A. acknowledges support from the Infrared Processing and Analysis Center (IPAC) Visiting Graduate Student Fellowship and from NASA/\textit{HST} grant SNAP-15922 from the Space Telescope Science Institute (STScI), which is operated by the Association of Universities for Research in Astronomy (AURA), Inc., under NASA contract NAS5-26555.
A.V.F. is grateful for financial
assistance from the Christopher R. Redlich Fund, the TABASGO Foundation,
and the U.C. Berkeley Miller Institute for Basic Research in Science (where he is a Senior Miller Fellow); additional funding was provided by NASA/{\it HST} grant AR-14295 from STScI.
G.F. acknowledges support from CONICET through grant PIP-2015-2017-11220150100746CO and from ANPCyT through grant PICT-2017-3133.

This paper made use of data from the Las Cumbres Observatory global network of telescopes through the Global Supernova Project.
Some of the observations reported herein were obtained at the Bok 2.3\,m telescope, a facility of the University of Arizona, at the MMT Observatory, a joint facility of the University of Arizona and the Smithsonian Institution, and at the W.~M.~Keck Observatory, which is operated as a scientific partnership among the California Institute of Technology, the University of California, and NASA; the Keck Observatory was made possible by the generous financial support of the W.~M.~Keck Foundation. 
This work is partly based on observations made with the NASA/ESA \textit{Hubble Space Telescope}, obtained from the Data Archive at STScI. These observations are associated with programs GO-9788, GO-13007, and GO-15151. Financial support for program GO-15151 was provided by NASA through a grant from STScI. 
This work is based in part on observations made with the \textit{Spitzer Space Telescope}, which is operated by the Jet Propulsion Laboratory, California Institute of Technology, under a contract with NASA. 
We thank the support of the staffs at the Neil Gehrels \textit{Swift} Observatory.
This research has made use of the NASA/IPAC Extragalactic Database (NED), which is funded by NASA and operated by the California Institute of Technology, as well as IRAF, which is distributed by NOAO (operated by AURA, Inc.), under cooperative agreement with NSF.
Numerical computations were in part carried out on the PC cluster at the Center for Computational Astrophysics, the National Astronomical Observatory of Japan.

The authors wish to recognise and acknowledge the very significant cultural role and reverence that the summits of Maunakea and Haleakal$\bar{\text{a}}$ have always had within the indigenous Hawaiian community. We are most fortunate to have the opportunity to conduct observations from these mountains.

\item[Author Contributions]
Daichi Hiramatsu initiated the study, triggered follow-up observations, reduced the Las Cumbres data, produced the light-curve models, performed the analysis, and wrote the manuscript.
D. Andrew Howell is the principal investigator of the Las Cumbres Observatory Global Supernova Project through which all of the Las Cumbres data were obtained; he also assisted with data interpretation and the manuscript.
Schuyler D. Van Dyk is the principal investigator of the \textit{HST} program `The Stellar Origins of Supernovae' (GO-15151) through which the post-explosion \textit{HST} data were obtained; he also found the progenitor candidate in the pre-explosion \textit{HST} F814W image, calculated the upper  limits in the pre-explosion \textit{HST} and \textit{Spitzer} images, and assisted with data interpretation and the manuscript.
Jared A. Goldberg produced the progenitor and light-curve models and assisted with their interpretation and the manuscript.
Keiichi Maeda assisted with theoretical nebular spectral model interpretation and the manuscript.
Takashi J. Moriya and Nozomu Tominaga assisted with theoretical SAGB progenitor and ECSN light-curve model interpretations and the manuscript.
Ken’ichi Nomoto assisted with theoretical SAGB progenitor and ECSN explosion model interpretation and the manuscript.
Griffin Hosseinzadeh assisted in obtaining the Las Cumbres data, reduced the FLOYDS spectra, and contributed comments to the manuscript.
Iair Arcavi, Curtis McCully, and Jamison Burke assisted in obtaining the Las Cumbres data;
Iair Arcavi and Curtis McCully also contributed comments to the manuscript.
K. Azalee Bostroem obtained the Keck LRIS and DEIMOS spectra, reduced the LRIS spectra, and contributed comments to the manuscript.
Stefano Valenti is the principal investigator of the Keck proposals (2018B, project code U009; 2019A, project code U019; 2019B, project code U034) under which the nebular spectra were obtained; he also built the Las Cumbres photometric and spectroscopic reduction pipelines, reduced the Keck DEIMOS spectrum, and contributed comments to the manuscript.
Yize Dong assisted in obtaining the Keck LRIS and DEIMOS spectra.
Peter J. Brown obtained and reduced the \textit{Swift} UVOT data.
Jennifer E. Andrews obtained and reduced the MMT and Bok spectra.
Christopher Bilinski reduced and analysed the MMT SPOL spectropolarimetry.
G. Grant Williams is the principal investigator of the Supernova Spectropolarimetry (SNSPOL) project.
Paul S. Smith is the principal investigator of the SPOL instrument.
G. Grant Williams and Paul S. Smith collected the spectropolarimetric data with the SPOL instrument at the MMT Observatory. 
Nathan Smith is the principal investigator of the MMT and Bok programs; he also contributed comments to the manuscript.
David J. Sand co-leads the University of Arizona team that obtained the MMT and Bok spectra; he also contributed comments to the manuscript.
Gagandeep S. Anand reduced and analysed the archival \textit{HST} WFC3/IR data (GO-12206) of the host galaxy NGC~2146.
Chengyuan Xu and Curtis McCully analysed and rejected the cosmic rays in the pre-explosion \textit{HST} F814W image.
Alexei V. Filippenko, Melina C. Bersten, Gast\'{o}n Folatelli, and Patrick L. Kelly are Co-Is of the \textit{HST} program (GO-15151); they also contributed comments to the manuscript (which Alexei V. Filippenko edited in detail).
Toshihide Noguchi monitored the supernova and provided his photometry.
Koichi Itagaki is the discoverer of the supernova; he also monitored the supernova and provided his photometry.
 
\item[Author Information] The authors declare that they have no competing financial interests. Correspondence and requests for materials should be addressed to D.~Hiramatsu (dhiramatsu@lco.global).


\end{addendum}


\clearpage

\newpage

\newpage

\newpage

\clearpage

\begin{methods}

\subsection{Extinction.}
\label{sec:extinction}

We obtained the Milky Way (MW) extinction\cite{Schlafly2011} of $A_{V,\rm{MW}}=0.258$\,mag via the NASA/Infrared Processing and Analysis Center (IPAC) Infrared Science Archive. 
We measure the total Na~{\sc i}~D EW of each host and MW component using the MMT Blue Channel (BC) spectra (moderate resolution of $1.45\,\Angstrom$) taken $3.6$--$53.4$\,d after the explosion (Extended Data Fig.~\ref{EDfig:allspec}). Since the ratio of the total Na~{\sc i}~D EW of the host to MW varies between $1.07$ and $1.25$, we estimate $A_{V,\rm{host}} \gtrsim A_{V,\rm{MW}}$.
As a cross-check, we transform the {\it gri} magnitudes of SN~2018zd to {\it VRI} magnitudes\cite{Jester2005} and compare the $V-I$ colour to that of well-observed, low-extinction Type~II-P SNe 1999em\cite{Leonard2002b}, 1999gi\cite{Leonard2002}, and 2017eaw\cite{VanDyk2019} by assuming $A_{V,\rm{host}} = A_{V,\rm{MW}}$ for SN~2018zd. Since the $V-I$ colour of SN~2018zd during the photospheric phase is consistent with the other SNe, we adopt a host extinction of $A_{V,\rm{host}} = A_{V,\rm{MW}}$ and assume a reddening law\cite{Fitzpatrick1999} with $R_V=3.1$. This extinction value is also consistent with the lower limit obtained from the spectropolarimetry (Supplementary Information). 
Increasing (or decreasing) the host extinction by more than $0.10$\,mag makes the $V-I$ colour inconsistent with that of the other Type~II-P SNe. Thus, we adopt a host extinction uncertainty of $\pm 0.10$\,mag.

\subsection{Luminosity distance.}
\label{sec:dL}

We apply the expanding photosphere method (EPM)\cite{Leonard2002,Dessart2005} and the standard candle method (SCM)\cite{Polshaw2015} using the measured Fe~{\sc ii} $\lambda$5169 velocities and transforming the {\it gri} to {\it VRI} magnitudes\cite{Jester2005}, which yields $6.5\pm0.7$ and $9.6\pm1.0$\,Mpc, respectively. 
The EPM is best used at early times ($\lesssim 30$\,d) when SN emission can be approximated as a diluted blackbody in free expansion\cite{Dessart2005}. However, the early emission from SN~2018zd is dominated by CSM interaction (Fig.~\ref{fig:uvUV} and Extended Data Fig.~\ref{EDfig:allvel}), making the EPM unreliable. 
On the other hand, the SCM is based on the luminosity--velocity correlation\cite{Hamuy2002,Kasen2009,Goldberg2019} at day 50 when the CSM interaction is negligible, which is well reproduced by our \texttt{MESA}+\texttt{STELLA} models (`\texttt{MESA}+\texttt{STELLA} progenitor and light-curve models' section and Extended Data Fig.~\ref{EDfig:LCscaling}). Thus, we favour the SCM over the EPM. 

It has been suggested that NGC 2146 may be farther away than the SCM estimate. There is a claim of a preliminary tip of the red giant branch (TRGB) distance obtained from archival \textit{HST} Wide Field Camera 3 IR channel (WFC3/IR) data (program GO-12206, principal investigator: M.~Westmoquette) that places the galaxy out at $\sim18$\,Mpc (ref.~\cite{Adamo2012}). We have independently reduced and analysed these same data and find that the single orbit of observations available (split between F110W and F160W) does not reach the necessary depths to make this conclusion. Even at the closer $10$\,Mpc distance, the archival data would not allow us to obtain a TRGB measurement owing to the short exposure times and intense levels of crowding. We also find that there are no archival \textit{HST} optical data of sufficient depths to obtain a TRGB measurement.

Future SN-independent distance measurements (for example, Cepheids and TRGB with \textit{HST}) will be necessary to verify the SCM estimate. We discuss the implications if the luminosity distance were larger than the SCM estimate in Supplementary Information.

\subsection{\textit{HST} and \textit{Spitzer} progenitor detection and upper limits.}
\label{sec:HST}

We were able to locate astrometrically the site of SN~2018zd in existing pre-explosion \textit{HST} archival images, specifically data obtained in bands F814W and F658N with the Advanced Camera for Surveys (ACS)/WFC instrument on 2004 April 10 (program GO-9788, principal investigator L.~Ho, with total exposure times of 120\,s and 700\,s, respectively; the F814W image consists of a single exposure), as well as in F225W with WFC3/UVIS on 2013 March 7 (program GO-13007, principal investigator L.~Armus; total exposure time of 1500\,s). We identified a potential candidate progenitor precisely by obtaining images of the SN itself on 2019 May 19 in F555W and F814W with WFC3/UVIS, as part of program GO-15151 (principal investigator S.~Van Dyk). We were able to astrometrically register the 2019 F814W image mosaic to the 2004 one using 23 stars in common, with a root-mean-square uncertainty of 0.14 ACS/WFC pixel. Furthermore, in a similar fashion we were able to match precisely the SN image with the F658N and F225W images as well; however, the progenitor candidate was not detected in either of those bands. We show the pre- and post-explosion images in Extended Data Fig.~\ref{EDfig:discovery}.

We extracted photometry from all of the \textit{HST} images using the package Dolphot\cite{Dolphin2016}. We found that Dolphot detected and measured a source at the position of the progenitor candidate with $m_{\rm F814W}=25.05 \pm 0.15$\,mag. Unfortunately, as noted above, the F814W pre-explosion observation consisted of only a single exposure, so it was not possible for the standard STScI pipeline to reject cosmic-ray hits in the usual way, while constructing an image mosaic from the single frame, as would normally be the case for two or more dithered exposures. In addition, we note that the flux measurement with Dolphot may be affected by the presence of cosmic-ray hits in the image at or around the progenitor site. Nevertheless, the values of both the Dolphot output parameters `object type' (1) and `sharpness' ($-0.013$) appear to point to the source being stellar-like. 

To determine whether the peak pixel seen at the candidate location is indeed merely a cosmic-ray hit or is the actual peak of a stellar point-spread function (PSF), we employ a deep-learning model (C.X. et al., manuscript in preparation) based on the results from DeepCR\cite{Zhang2020DeepCR}. We find that no pixels in the vicinity of the candidate progenitor have a model score higher than $5.1\times 10^{-5}$. If we use this score as a threshold, the model has a completeness of $99.93\%$ based on the test data taken with the same instrument. We therefore conclude that progenitor candidate is a real PSF with $>3\sigma$ confidence. If the object was not actually detected, we find that the upper limit at 3$\sigma$ to detection in F814W is $>26.3$ mag. 

Inserting and recovering an artificial star of varying brightness at the exact SN position with Dolphot in both F225W and F658N led to estimates of the upper limits to detection (at 3$\sigma$) of $>23.6$ and $>24.1$\,mag, respectively. In addition, note that we measured a brightness of the SN itself in the 2019 \textit{HST} observations of $m_{\rm F555W}=21.53 \pm 0.01$ and $m_{\rm F814W}=20.33 \pm 0.01$\,mag.

The SN site also can be found in pre-explosion \textit{Spitzer} data both from the cryogenic and so-called warm (post-cryogenic) missions, from 3.6\,$\mu$m to 24\,$\mu$m. The data are from observations with the Infrared Array Camera (IRAC\cite{Fazio2004}) in channels 1 (3.6\,$\mu$m) and 2 (4.5\,$\mu$m; the SN site sits in a gap of spatial coverage in channels 3 and 4 at 5.8\,$\mu$m and 8\,$\mu$m, respectively) on 2004 March 8 (program 59, principal investigator G.~Rieke) and on 2007 October 16 (program 40410, principal investigator G.~Rieke) in channels 2 and 4; from observations with the Multiband Imaging Photometer for Spitzer (MIPS\cite{Rieke2004}) at 24\,$\mu$m on 2004 March 16 (program 59, principal investigator G.~Rieke; the sensitivity and resolution of the data at 70\,$\mu$m and 160\,$\mu$m are not sufficient to hope to detect the progenitor and were not considered further); and from observations with IRAC in channels 1 and 2 on 2011 November 15 (program 80089, principal investigator D.~Sanders). We show the 2011 November 15 IRAC observation in channel 1 in Extended Data Fig.~\ref{EDfig:discovery}.

Observations with IRAC of the SN itself were obtained on 2019 January 24 (program 14098, principal investigator O.~Fox); however, we did not analyse these data, other than to extract an absolute position for the SN of $\alpha = 6^{\rm h}18^{\rm m}03.43^{\rm s}$, $\delta = +78^\circ 22' 01{\farcs}4$ (J2000; $\pm 0{\farcs}3$ in each coordinate). Using MOPEX\cite{Makovoz2006} we constructed mosaics from all of the useful pre-SN imaging data, and with APEX within MOPEX\cite{Makovoz2005} we inserted into the images an artificial star of varying brightness at this absolute position. From this, we estimated upper limits to detection of the progenitor (at 3$\sigma$) of $>19.0$ and $>18.1$\,mag in channels 1 and 2 (respectively) from the 2004 March 8 data; $>18.1$ and $>14.5$\,mag in channels 2 and 4 (respectively) from 2007 October 16; and $>19.0$ and $>18.4$\,mag in channels 1 and 2 (respectively) from 2011 November 15 (we have assumed the zeropoints from the IRAC Instrument Handbook). We also estimated $>10.2$\,mag at 24\,$\mu$m from the 2004 March 16 observation (we have assumed the zeropoint from the MIPS Instrument Handbook).

We show the resulting spectral energy distribution (SED), or limits thereon, for the SN~2018zd progenitor in Extended Data Fig.~\ref{EDfig:progenitor_sed}. The distance ($9.6\pm1.0$\,Mpc) and extinction ($A_V=0.52\pm0.10$\,mag) to the SN were adopted (`Luminosity distance' and `Extinction' sections), assuming that the latter also applied to the progenitor as well. We have further assumed a reddening law\cite{Fitzpatrick1999} with $R_V=3.1$ and extended it into the mid-infrared\cite{Xue2016}.
For comparison, we also show single-star SAGB and RSG (with respective initial masses $M_{\rm init} = 8$ and $15\,\Msun$) models from BPASS v2.2\cite{Stanway2018} with metallicities $Z=0.020$ (solar) and $Z=0.010$ (subsolar; as discussed in Supplementary Information, the SN site metallicity is probably subsolar). We have further included for comparison the observed SED for the candidate SAGB star MSX SMC 055 (IRAS 00483$-$7347)\cite{Groenewegen2018} as well as the SED for the progenitor of the low-luminosity Type~II-P SN~2005cs\cite{Maund2005,Li2006}. 

We note that the SEDs for the BPASS RSG models with $M_{\rm init}=15\,\Msun$ are probably not realistic, since they are merely bare photospheres, whereas we would expect such a star to possess a dusty CSM, as was the case for the progenitor of SN~2017eaw\cite{VanDyk2019}. The same could also potentially be said for the SAGB models, given the dusty nature of MSX SMC 055. Again, these BPASS model SEDs are bare photospheres and do not include CSM, for the presence of which we have strong evidence (given here) in the case of the SN~2018zd progenitor; this merits further development of the SED models including the effect of dusty CSM. 

It is difficult to infer much about the nature of the SN~2018zd progenitor, based on a probable detection in one band and upper limits in the others. However, its inferred SED does appear to be less consistent with that of an $M_{\rm init} \gtrsim 8\,\Msun$ RSG star, as well as the SN~2005cs progenitor, and more consistent with a potentially dusty SAGB star, such as MSX SMC 055. If there were circumstellar dust around the SN~2018zd progenitor, it was destroyed as the SN shock progressed through.

We should revisit this site either with \textit{HST} or the \textit{James Webb Space Telescope} in a bandpass similar to F814W, when the SN has sufficiently faded, to confirm that the candidate object was indeed the progenitor. Again, we cannot entirely rule out that the source detected in the pre-SN image at the precise SN position is not related to a cosmic-ray hit; however, all of the indications suggest this is a real detected star, which should have vanished when the SN site is observed at a sufficiently late time.

\subsection{\texttt{MESA}+\texttt{STELLA} progenitor and light-curve models.}
\label{sec:MESAmod}

Recent work\cite{Goldberg2019, Dessart2019, Bersten2011, Martinez2019} has highlighted the non-uniqueness of bolometric light-curve modeling for extracting explosion characteristics (ejecta mass $M_\mathrm{ej}$, explosion energy $E_\mathrm{exp}$, and progenitor radius $R$) from plateau features (in particular, luminosity at day 50, $L_\mathrm{50}$, and plateau duration, $t_\mathrm{p}$) without an independent prior on one of $M_\mathrm{ej}$, $E_\mathrm{exp}$, or $R$. 
Owing to the presumed presence of dense CSM and its potential influence on the early light curves and velocities, shock-cooling modeling and early expansion velocities cannot simply lift this degeneracy.

To allow light-curve analysis to be agnostic to the progenitor mass, three different explosion models were created with equally good by-eye matches to the bolometric light curve and expansion-velocity data on the plateau. The progenitor models were selected from a pre-existing grid\cite{Goldberg2020} of \texttt{MESA}\cite{Paxton2018,Paxton2019} RSG progenitor models with expected ejecta masses and radii within the family of explosions consistent with the $L_{50}$, $t_\mathrm{p}$, and $M_\mathrm{Ni}$ of SN~2018zd (Extended Data Fig.~\ref{EDfig:LCscaling}; see Supplementary Information for the model details). 

The explosion energies for each model were then chosen and adjusted to match the light curve of SN~2018zd with the respective progenitor model radii using the degeneracy relations\cite{Goldberg2019}
\begin{equation}
\begin{split}
\begin{aligned}
\log(E_{51}) &= -0.728 + 2.148 \log(L_\mathrm{p,42}) - 0.280\log(M_\mathrm{Ni}) + 2.091\log(t_\mathrm{p,2}) - 
1.632\log(R_{500}), \\
\log(M_{10}) &= -0.947 + 1.474 \log(L_\mathrm{p,42}) - 0.518 \log(M_\mathrm{Ni}) + 3.867 \log(t_\mathrm{p,2}) - 1.120 \log(R_{500}),
\label{eq:EandMofR}
\end{aligned}
\end{split}
\end{equation}
where $E_{51} = E_\mathrm{exp}/10^{51}$\,erg, $M_{10}=M_\mathrm{ej}/10\,M_\odot$, $M_\mathrm{Ni}$ is in units of $M_\odot$, $L_\mathrm{p,42} = L_{50}/10^{42}$\,erg\,s$^{-1}$, $t_\mathrm{p,2}=t_\mathrm{p}/100\,\mathrm{d}$, and $R_{500} = R/500\,R_\odot$. 
Plugging in $L_\mathrm{50} = 8.6\times10^{41}$\,erg\,s$^{-1}$ from the bolometric light curve at day 50, $t_\mathrm{p} = 125.4$\,d determined by fitting the drop from the plateau\cite{Valenti2016}, and observed $M_\mathrm{Ni}=0.0086\,M_\odot$, these relations describe the possible explosion parameter space (Extended Data Fig.~\ref{EDfig:LCscaling}).
They are intended for Ni-rich ($M_\mathrm{Ni}>0.03\,M_\odot$) Type~II-P SNe of RSG progenitors with no fallback, but nonetheless provide a heuristic estimate for the degeneracy between explosion energy, progenitor radius, and ejected mass.

This degeneracy motivates the set of progenitor models and explosion energies that we use to reproduce the light-curve properties, and reveals low recovered $E_\mathrm{exp}$ which overlap substantially with the expected parameter space of ECSNe. 
The mapping between $M_\mathrm{ej}$ recovered for Fe CCSNe and ECSNe is less robust, as differences in mixing extent and H/He abundances could account for differences in the recovered $M_\mathrm{ej}$ from explosions of different stellar progenitors\cite{Kasen2009,Kozyreva2019,Goldberg2019}. 
As seen in Extended Data Fig.~\ref{EDfig:LCscaling}, even though SN~2018zd is not particularly dim, low-energy explosions of radially extended progenitors can match the plateau luminosity. A slightly lower-$M_{\rm ej}$ progenitor with a radius of $1$,$400\,R_\odot$, for example, could even produce this luminosity with an explosion energy of $\sim1.5\times10^{50}$\,erg.

The explosions were carried out using \texttt{MESA} until near shock breakout. The models were then handed off to \texttt{STELLA}\cite{Blinnikov1998,Blinnikov2000,Blinnikov2006} to produce synthetic bolometric light curves and expansion velocities (Extended Data Fig.~\ref{EDfig:LCscaling}; see Supplementary Information for the modeling details). 
We see good agreement between all three models and observations (varying by at least 50\% in $M_\mathrm{ej}$, $E_\mathrm{exp}$, and $R$), with deviations at early times that can be attributed to the extended stellar atmosphere and potential interaction with the circumstellar environment.  

To account for the early deviations, we affix a wind-density profile with $\rho_\mathrm{wind}(r) = \dot{M}_\mathrm{wind}/4\pi r^2 v_\mathrm{wind}$, where $r$ is the radial extent, $\dot{M}_\mathrm{wind}$ is a constant wind mass-loss rate, and $v_\mathrm{wind}$ is the wind velocity for time $t_\mathrm{wind}$ (that is, $M_\mathrm{wind} = \dot{M}_\mathrm{wind} t_\mathrm{wind}$), onto the \texttt{MESA} model at handoff to \texttt{STELLA}. 
We construct a grid of CSM models by varying the following parameters: $\dot{M}_\mathrm{wind} = \{10^{-4},\,3\times10^{-4}, 10^{-3},\,3\times10^{-3},\,10^{-2},\,3\times10^{-2},\,10^{-1},\,3\times10^{-1}\}\,M_\odot$\,yr$^{-1}$ and $t_\mathrm{wind} =\{1,\,3,\,10,\,30\}$\,yr for each \texttt{MESA} model, assuming a typical wind velocity\cite{Moriya2014} $v_\mathrm{wind}=20$\,km\,s$^{-1}$. Then we perform $\chi^2$ fitting on the observed bolometric light curve over the full temporal evolution and find the best-fit parameters $\dot{M}_\mathrm{wind}=0.01\,M_\odot$\,yr$^{-1}$ and $t_\mathrm{wind}=10$\,yr. 

Remarkably, the best-fit parameters are the same for all three degenerate models, and also reproduce the early blueward UV-colour evolution (Extended Data Fig.~\ref{EDfig:LCscaling}). Thus, we choose model M8.3\_R1035\_E0.23 (Supplementary Information) as being representative and present it for the UV-colour plot in Fig.~\ref{fig:uvUV}. In addition to matching the early-time luminosity excess, a dense wind profile suppresses the early photospheric and Fe line velocities\cite{Moriya2018}. The kink seen in the modelled Fe line velocity with Sobolev optical depth $\tau_\mathrm{Sob}=1$ in the \texttt{STELLA} models can be attributed to numerics at the boundary between the CSM profile and the surface of the stellar ejecta (Extended Data Fig.~\ref{EDfig:LCscaling}). Overall, the models still yield general agreement between the calculated velocity evolution and the data.

We note that at $3$--$10$\,d after the explosion, the blackbody temperatures ($\sim20$,$000$--$25$,$000$\,K) may be underestimated (Supplementary Information), which could affect the luminosity around the peak, and so the CSM models as well. For the flash spectral model comparisons in Fig.~\ref{fig:flash}, we use a conservative temperature constraint of $\sim20$,$000$--$30$,$000$\,K.

\subsection{The rate of ECSNe.}
\label{sec:rate}

Among other previously suggested ECSN candidates (Supplementary Information and Extended Data Fig.~\ref{EDfig:ECSNcan}), Type~IIn-P SNe share similar properties to SN~2018zd. Thus, the Type~IIn-P SN rate may be related to the ECSN rate. As there is no rate estimation for Type~IIn-P SNe in the literature to our knowledge, we put a rough lower limit using publicly announced Type~IIn SNe on the Weizmann Interactive Supernova Data Repository (WISeREP)\cite{Yaron2012} and/or the Transient Name Server (TNS) by cross-checking with the literature and the Open Supernova Catalog\cite{Guillochon2017}, and also cross-correlating the public spectra to SN spectral libraries Superfit\cite{Howell2005} and SNID\cite{Blondin2007} when available. There are 528 objects classified as Type~IIn SNe on WISeREP and/or TNS (as of 2020 March~11). We exclude 73 objects as misclassified early-flash Type~II SNe, Type~Ia-CSM SNe, Type~Ibn SNe, SN imposters, or active galactic nuclei. Although 241 objects do not have enough public and/or published spectra and/or light curves to secure the Type~IIn classifications and/or to be identified as Type~IIn-P SNe, we include them in the further analysis so as not to overestimate the lower limit when taking a number ratio of Type~IIn-P to Type~IIn SNe (see below and Supplementary Fig.~\ref{SIfig:IIn-P}).

To identify Type~IIn-P SN candidates from the 455 objects, we apply two light-curve criteria based on the known Type~IIn-P SN characteristics: (1) the $V$, $r/R$, or $i/I$-band decline of less than $2$\,mag in the first $50$\,d after the explosion; and (2) the $V$, $r/R$, or $i/I$-band drop of more than $2$\,mag in $30$\,d within 100--150\,d after the explosion. This yields four Type~IIn-P SN candidates: SNe 2005cl ($z=0.025878$)\cite{Kiewe2012}, 2005db ($z=0.015124$)\cite{Kiewe2012}, 2006bo ($z=0.0153$)\cite{Taddia2013}, and 2011A ($z=0.008916$)\cite{deJaeger2015}. In addition, there are three known Type~IIn-P SNe: 1994W ($z=0.004116$)\cite{Sollerman1998}, 2009kn ($z=0.015798$)\cite{Kankare2012}, and 2011ht ($z=0.003646$)\cite{Mauerhan2013} (Supplementary Fig.~\ref{SIfig:IIn-P}).

To compare with the volume-limited ($\leq60$\,Mpc) Lick Observatory Supernova Search (LOSS) sample\cite{Smith2011}, we apply the same distance cut, leaving 42 Type~IIn SNe (17 and 25 with sufficient and insufficient data, respectively) and 3 Type~IIn-P SNe (SNe 1994W, 2011A, and 2011ht). As these SNe were discovered by different surveys with different strategies, we have no handle on the incompleteness (but also see Supplementary Fig.~\ref{SIfig:IIn-P}). Thus, we neglect the incompleteness and take the number ratio of Type~IIn-P to Type~IIn SNe within $60$\,Mpc multiplied by the LOSS Type~IIn SN rate\cite{Smith2011}, $3/42\times8.8\%=0.63\%$ of all CCSNe, as a rough lower limit of the Type~IIn-P SN rate.

The identification of SN~2018zd-like SNe from Type~II-P SNe ($48.2\%$ of all CCSNe) requires not only the light-curve morphology, but also the early UV colours and the flash and nebular spectroscopy, all of which combined are rarely available on WISeREP, TNS, and/or the Open Supernova Catalog. This current sample limitation may indicate that many SN~2018zd-like SNe have been overlooked as normal Type~II-P SNe. 
Given the limitation, we simply take the lowest possible limit of $>0$\% with the one identification of SN~2018zd as an ECSN.

By combining the estimated Type~IIn-P and SN~2018zd-like lower limits, we obtain a Type~IIn-P + SN~2018zd-like lower limit of $>0.6$\% of all CCSNe. 
From the nucleosynthetic point of view, ECSNe are expected to constitute $\lesssim8.5\%$ of all CCSNe\cite{Wanajo2018}. With the above estimates, the ECSN rate can be roughly constrained within $0.6$--$8.5$\% of all CCSNe, which corresponds to a narrow SAGB progenitor mass window of $\Delta M_{\rm SAGB} \approx 0.06$--$0.69$\,$M_\odot$ assuming maximum and minimum SAGB masses of $9.25\,M_\odot$ and $9.25\,M_\odot - \Delta M_{\rm SAGB}$ (respectively) at solar metallicity\cite{Poelarends2008} (Extended Data Fig.~\ref{EDfig:rate}).
We note that the Type~IIn-P and SN~2018zd-like rates are probably metallicity dependent (as is the SAGB mass window), but we defer more detailed analysis with a homogeneous sample in the future when one becomes publicly available.

\end{methods}

%



 
\begin{addendum}

\item[Data Availability] The data that support the plots within this paper and other findings of this study are available from the Open Supernova Catalog (\url{https://sne.space/}) and the Weizmann Interactive Supernova Data Repository (\url{https://wiserep.weizmann.ac.il/}), or from the corresponding author upon reasonable request.

\item[Code Availability] \texttt{MESA} is publicly available at \url{http://mesa.sourceforge.net/}.

\end{addendum}


\clearpage

\begin{addendum}

\item[Extended Data]


\begin{EDfigure}
\centering
\includegraphics[width=0.99\textwidth]{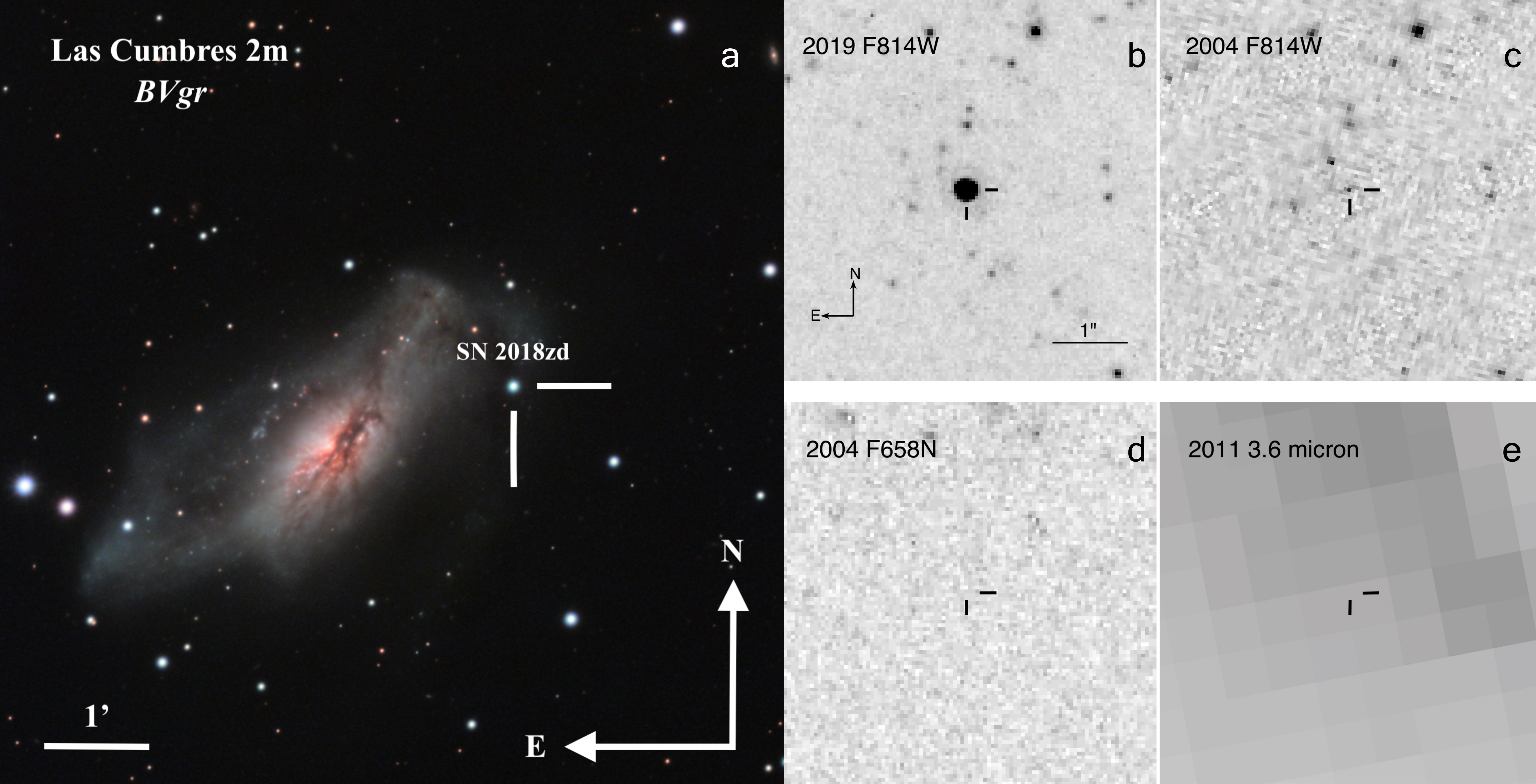}
\caption{
\textbf{The host galaxy and post- and pre-explosion images of SN~2018zd.}
\textbf{a,} Las Cumbres 2\,m {\it BVgr}-composite image of SN~2018zd and the host starburst galaxy NGC 2146 (Supplementary Information), courtesy of Peter~Il\'a\v{s}. 
At the assumed luminosity distance of $9.6$\,Mpc, $1'$ corresponds to $2.8$ kpc. 
SN~2018zd is on a tidal stream which was probably ejected during a galaxy merger event.
\textbf{b,} Portion of an \textit{HST} WFC3/UVIS F814W mosaic obtained on 2019 May 19, 443.7\,d after the explosion of SN~2018zd (indicated by the tick marks). 
\textbf{c,} Portion of an \textit{HST} ACS/WFC F814W mosaic from 2004 April 10; the SN site is similarly indicated by tick marks. 
This mosaic consists of a single exposure, so to remove a number of cosmic-ray hits in the image, we use a masked mean filter to smooth any pixels that have a score of 0.001 or higher from our deep-learning model (Methods). The pixels associated with the progenitor candidate had scores $<4\times10^{-5}$, so are not affected.
\textbf{d,} Same as panel (\textbf{c}), but with F658N on the same epoch. 
\textbf{e,} Portion of a \textit{Spitzer} IRAC 3.6\,$\mu$m mosaic obtained on 2011 November 15, with the SN site again indicated by tick marks. 
All panels (\textbf{b})--(\textbf{e}) are shown to the same scale and orientation, with north up and east to the left. 
The progenitor candidate is identified only in the single \textit{HST} ACS/WFC F814W image (\textbf{c}).
\label{EDfig:discovery}
}
\end{EDfigure}

\newpage

\begin{EDfigure}
\centering
\includegraphics[width=0.68\textwidth]{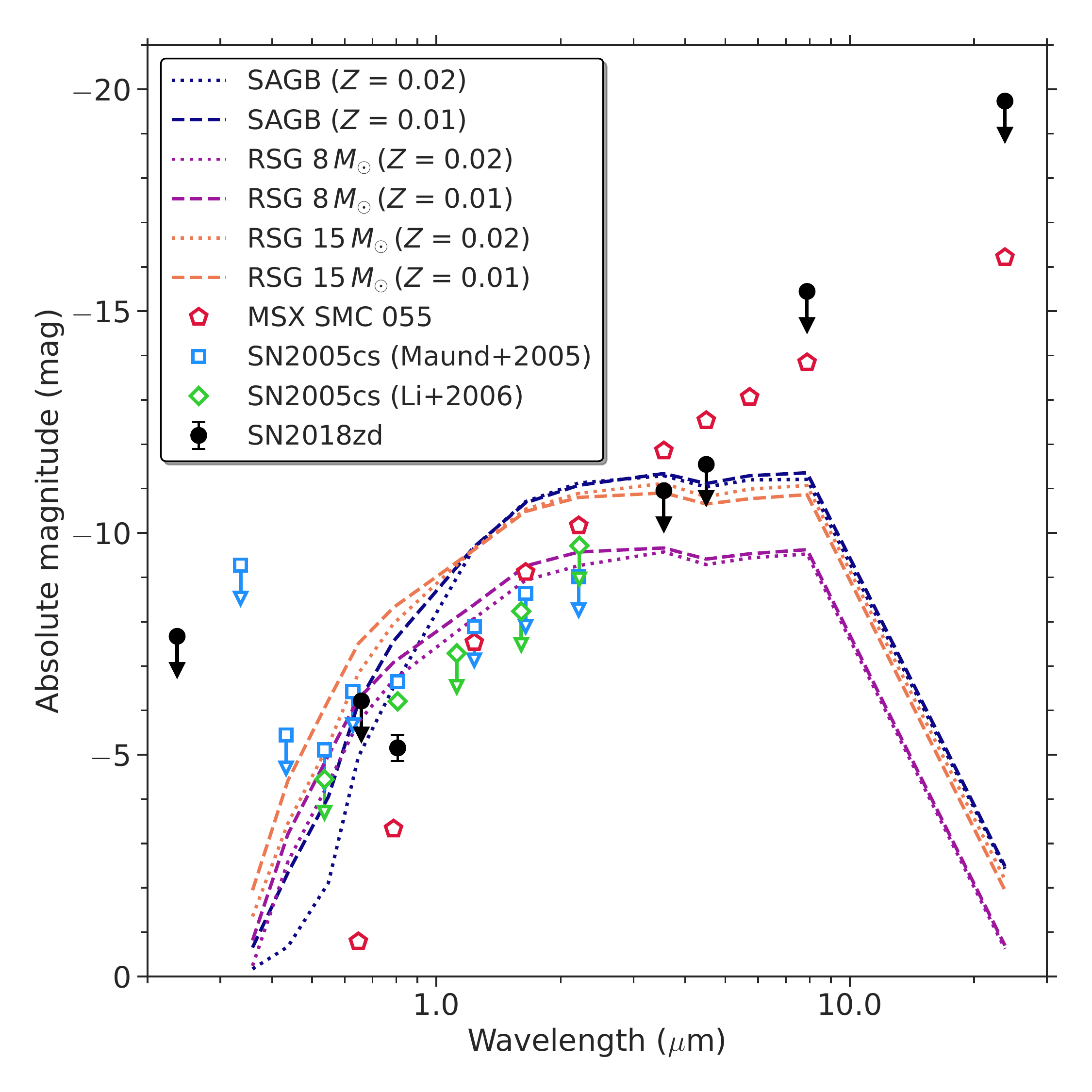}
\caption{
\textbf{SN progenitor and SAGB candidate SEDs.}
The SED for the SN~2018zd progenitor candidate resulting from pre-explosion \textit{HST} and \textit{Spitzer} archival data (Methods; black solid circles). For comparison we show model SEDs from BPASS v2.2\cite{Stanway2018} for SAGB stars (in the initial mass range $M_{\rm init}=6$--$8\,\Msun$ with bolometric luminosities $L\approx 10^5\,\Lsun$ in the last model timestep; navy curves) and RSG stars at $M_\mathrm{init}=8\,\Msun$ (purple curves) and $M_\mathrm{init}=15\,\Msun$ (orange curves), at metallicities $Z=0.02$ (solar; short-dashed line) and $Z=0.01$ (subsolar; long-dashed line).
The SEDs of the BPASS models are extrapolated into the mid-infrared via MARCS\cite{Gustafsson2008} model stellar atmospheres of similar temperatures as the last BPASS model timesteps, deriving synthetic photometry from those atmosphere models using the bandpass throughputs provided in the \textit{Spitzer} IRAC and MIPS Instrument Handbooks. 
Also shown for comparison are the SEDs for the SAGB candidate MSX SMC 055 (assuming Galactic foreground extinction and adjusted to a Small Magellanic Cloud distance modulus of $\mu=18.90$\,mag from the Extragalactic Distance Database\cite{Tully2009}; red open pentagons\cite{Groenewegen2018}) and for the progenitor of the low-luminosity Type~II-P SN~2005cs (assuming the total reddening from the two studies\cite{Maund2005,Li2006} and adjusted to a recent accurate distance for M51\cite{McQuinn2016}; blue open squares\cite{Maund2005}, green open diamonds\cite{Li2006}). The luminosity of the \textit{HST} ACS/WFC F814W detection of the SN~2018zd progenitor candidate lies between MSX SMC 055 and the SN~2005cs progenitor.
\label{EDfig:progenitor_sed}
}
\end{EDfigure}

\newpage

\begin{EDfigure}
\centering
\includegraphics[width=0.99\textwidth]{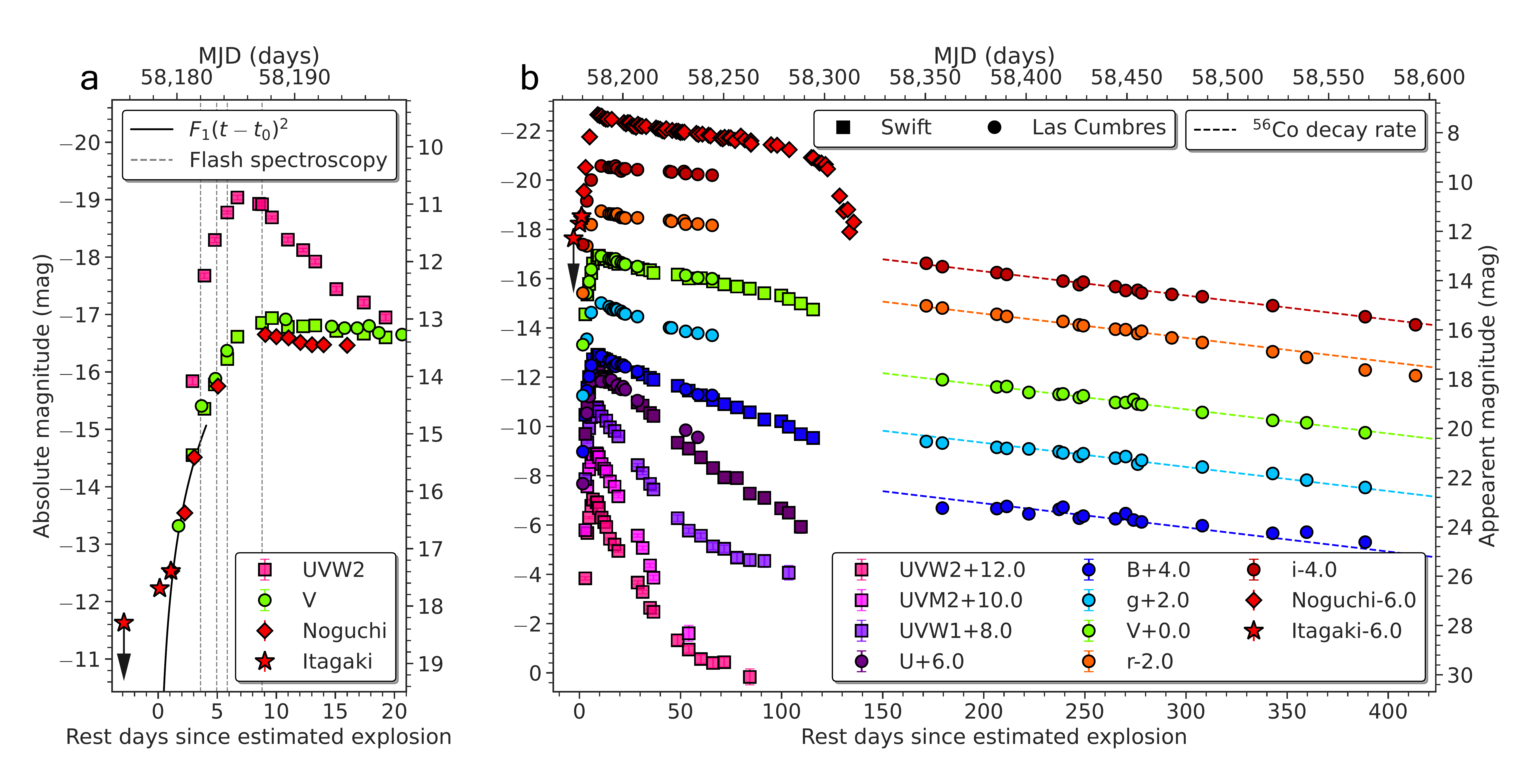}
\caption{
\textbf{Multiband light curve of SN~2018zd.}
\textbf{a,} Multiband light curve of SN~2018zd focusing on the early rise. 
A quadratic function $F_1(t-t_0)^2$ is fitted to the unfiltered optical Itagaki and the first three Noguchi points to estimate an explosion epoch $t_0=$\,MJD\,$58178.4\pm0.1$ (Supplementary Information). The observed flash-spectroscopy epochs (Extended Data Fig.~\ref{EDfig:allspec}) are marked by the vertical dashed lines. Note the sharper rise in the \textit{Swift} $UVW2$ than in the $V$ and unfiltered photometry during the flash-spectroscopy epochs. 
\textbf{b,} Multiband light curve of SN~2018zd up to the $^{56}$Co decay tail. 
The data gap is due to the Sun constraint. 
Error bars denote $1\sigma$ uncertainties and are sometimes smaller than the marker size. 
The light-curve shape resembles that of a typical Type~II-P SN. Comparing the luminosity on the tail to that of SN~1987A\cite{Hamuy2003}, we estimate a $^{56}$Ni mass of $(8.6\pm0.5)\times10^{-3}\,M_\odot$ at the assumed luminosity distance of $9.6$\,Mpc.
\label{EDfig:allLC}
}
\end{EDfigure}

\newpage

\begin{EDfigure}
\centering
\includegraphics[width=0.68\textwidth]{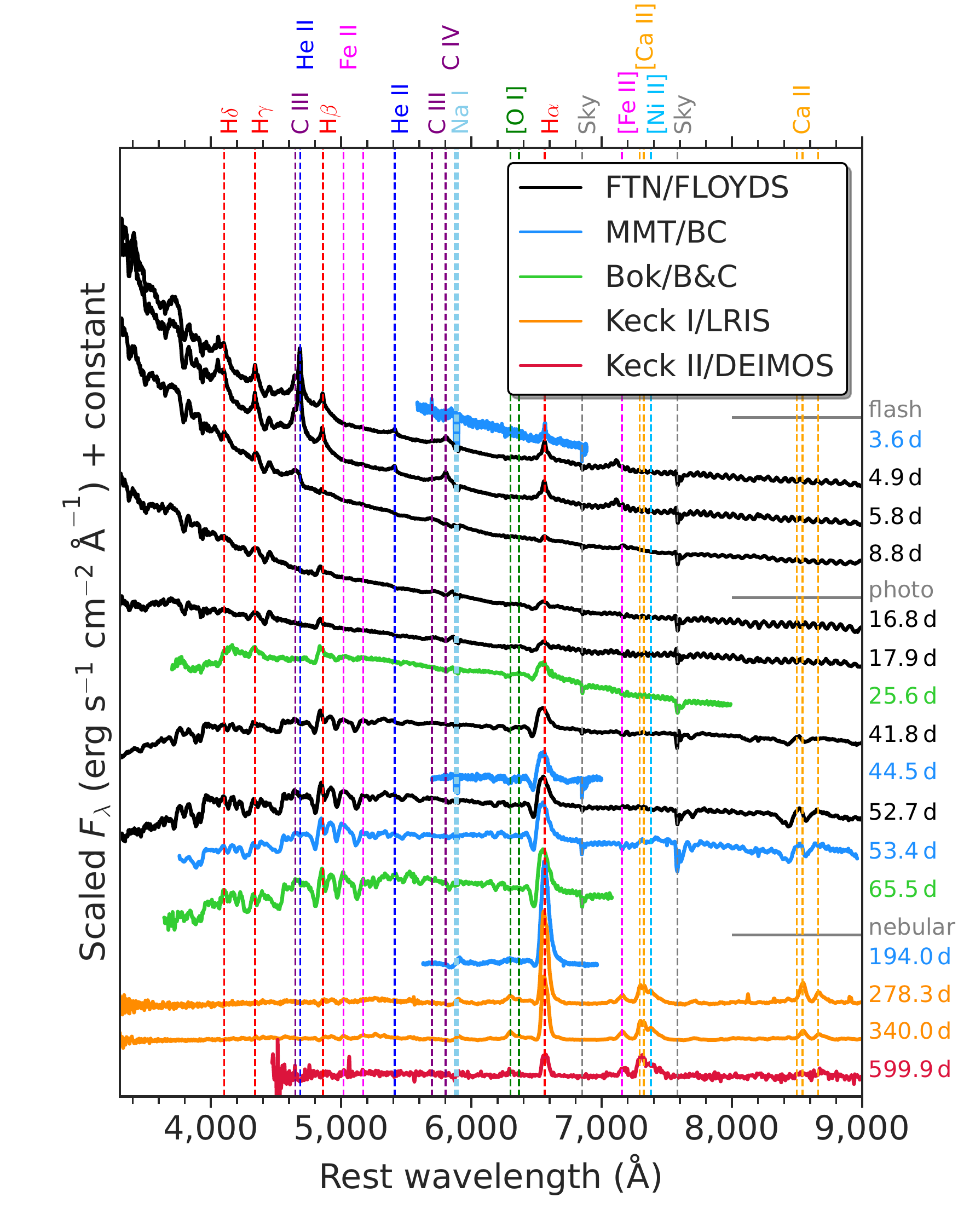}
\caption{
\textbf{Optical spectral time series of SN~2018zd.}
The flash features (for example, He~{\sc ii}, C~{\sc iii}, and C~{\sc iv}) persist up to $>8.8$\,d and disappear before $16.8$\,d. 
Then the broad Balmer-series P~Cygni lines appear, typical of the photospheric phase of a Type~II-P SN. 
After $\sim200$\,d, the nebular emission lines (for example, H$\alpha$, [Ca~{\sc ii}], and [Ni~{\sc ii}]) dominate over the relatively flat continuum.
\label{EDfig:allspec}
}
\end{EDfigure}

\newpage

\begin{EDfigure}
\centering
\includegraphics[width=0.99\textwidth]{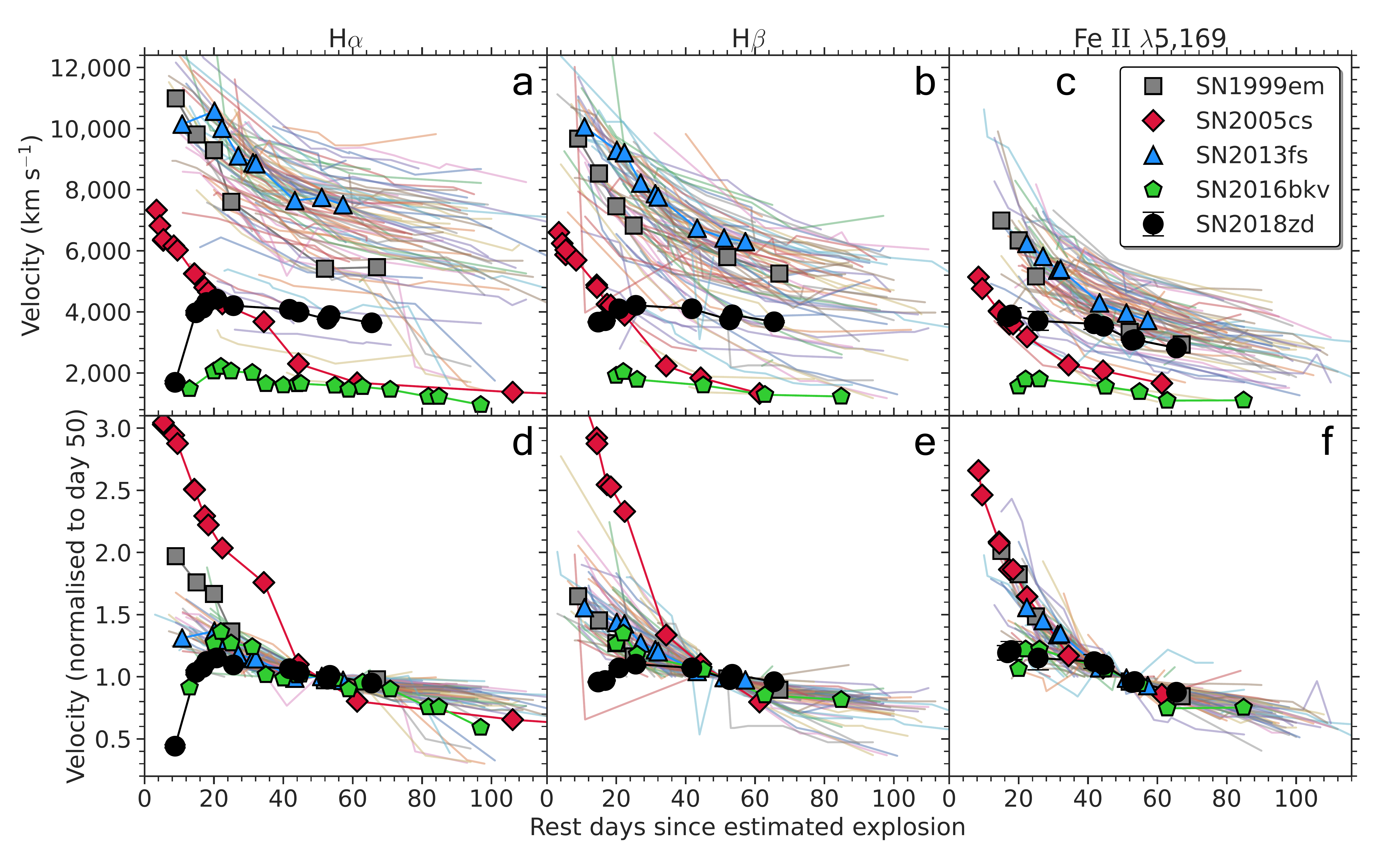}
\caption{
\textbf{Expansion velocities as a function of time.}
Comparison of the unnormalised (\textbf{a, b, c}) and normalised (to day 50; \textbf{d, e, f}) H$\alpha$, H$\beta$, and Fe~{\sc ii} $\lambda$5169 expansion velocities of SN~2018zd (Supplementary Information) with a Type~II SN sample\cite{Gutierrez2017} (transparent lines), including archetypal SN~1999em, along with low-luminosity SN~2005cs\cite{Pastorello2009}, early-flash SN~2013fs\cite{Yaron2017}, and low-luminosity and early-flash SN~2016bkv\cite{Hosseinzadeh2018,Nakaoka2018}. 
Error bars denote $1\sigma$ uncertainties and are sometimes smaller than the marker size. 
Note the pronounced early H$\alpha$ and H$\beta$ rises and the relatively flat velocity evolution (up to $\sim30$\,d) of SN~2018zd, indicating shock propagation inside the dense, optically-thick CSM. 
\label{EDfig:allvel}
}
\end{EDfigure}

\newpage

\begin{EDfigure}
\centering
\includegraphics[width=0.99\textwidth]{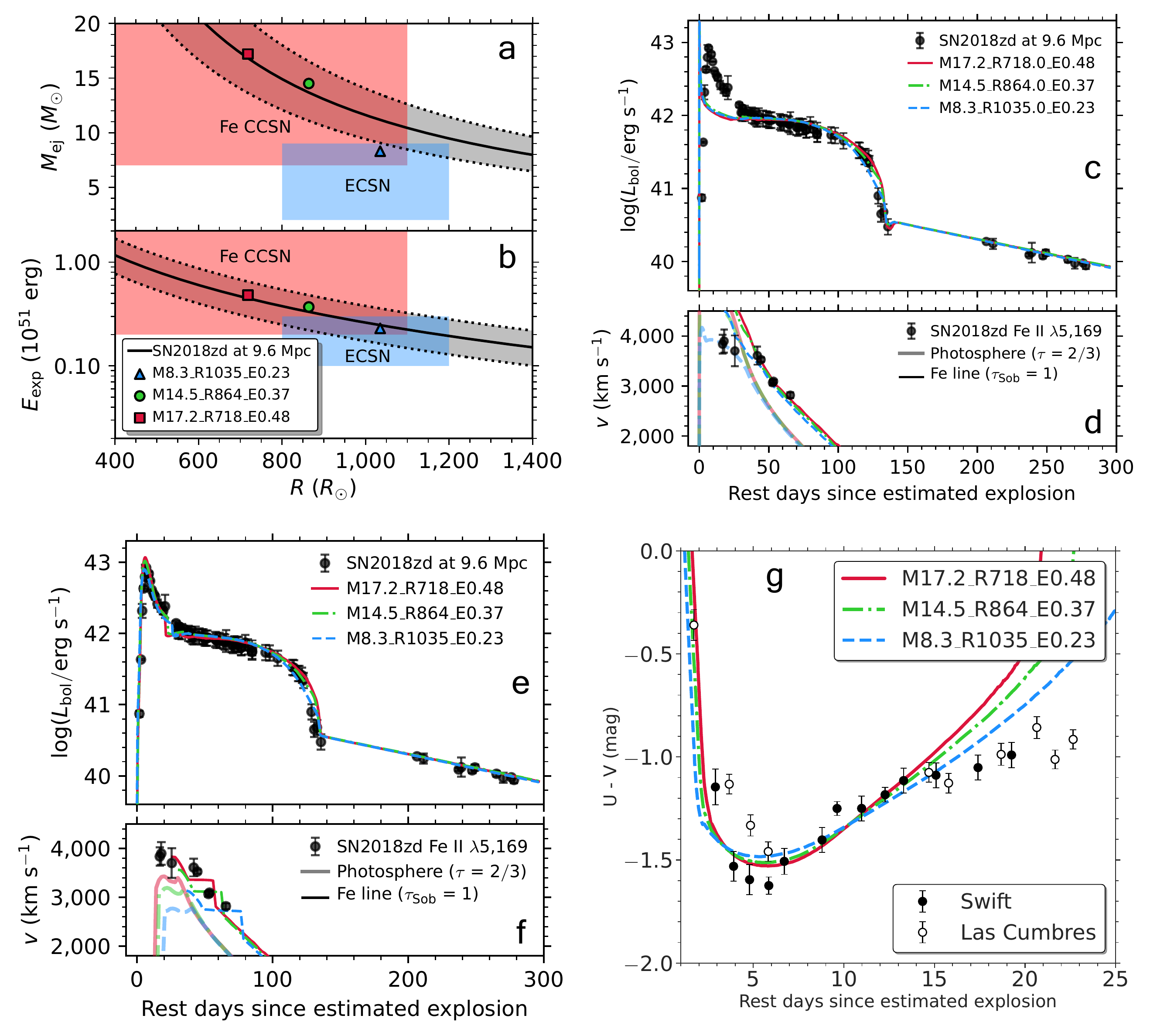}
\caption{
\textbf{\texttt{MESA}+\texttt{STELLA} progenitor and degenerate light-curve models.}
\textbf{a, b,} Ejecta mass $M_\mathrm{ej}$ and explosion energy $E_\mathrm{exp}$ inferred from Eq.~(\ref{eq:EandMofR}) (Methods) as a function of progenitor radius $R$ consistent with the bolometric light curve  of SN~2018zd at the assumed luminosity distance of $9.6\pm1.0$\,Mpc, along with the properties of the three degenerate explosion models. 
The blue and red shaded regions show explosion parameters expected for ECSNe\cite{Tominaga2013,Jones2013,Doherty2017} and typical of Fe CCSNe\cite{Sukhbold2016}, respectively.
\textbf{c, d,} Three degenerate \texttt{MESA+STELLA} explosion models providing good fits to the light curve and velocities inferred from the Fe~{\sc ii} $\lambda$5169 line during the plateau phase. 
Models are labeled by $\mathrm{M}[M_\mathrm{ej,\odot}]\_\mathrm{R}[R_\odot]\_\mathrm{E}[E_\mathrm{exp,51}]$. 
Error bars denote $1\sigma$ uncertainties. 
Note the observed early-time excess luminosity and suppressed velocity of SN~2018zd. This light-curve degeneracy highlights the inability to distinguish ECSNe from Fe CCSNe solely based on their light curves, suggesting that many ECSNe might have been overlooked owing to the lack of additional observations.
\textbf{e, f,} Same as panels (\textbf{c, d}), but adding a dense wind profile ($\dot{M}_\mathrm{wind}=0.01\,M_\odot$\,yr$^{-1}$, $v_\mathrm{wind}=20$\,km\,s$^{-1}$, and $t_\mathrm{wind}=10$\,yr) to the three degenerate \texttt{MESA} models before handoff to \texttt{STELLA}. 
\textbf{g,} Comparison of the UV-colour models with the same wind CSM parameters as in panels (\textbf{e, f}). 
Error bars denote $1\sigma$ uncertainties. 
All three models with the same wind CSM parameters are able to reproduce the early-time luminosity excess and blueward UV colour evolution almost identically, suggesting the insensitivity of a particular model choice. 
Despite a possible artificial velocity kink when the Fe line-forming region transitions from the CSM to the stellar ejecta, the velocity evolution with the early suppression is also reproduced.
\label{EDfig:LCscaling}}
\end{EDfigure}

\newpage

\begin{EDfigure}
\centering
\includegraphics[width=0.99\textwidth]{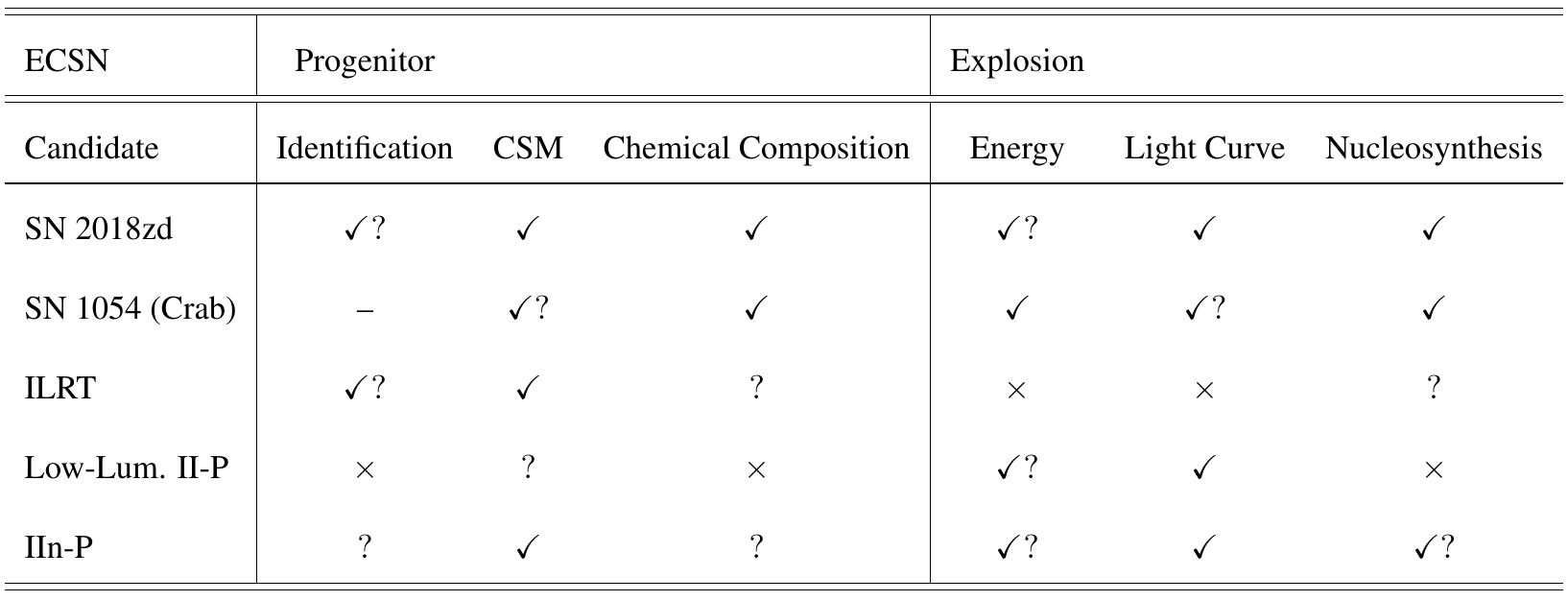}
\caption{
\textbf{ECSN candidate checklist.} 
Check marks, check+question marks, and cross marks (respectively) indicate observations consistent, perhaps consistent, and inconsistent with theoretical expectations. Dashed lines indicate the lack of observational constraints, and lone question marks indicate unknowns (Supplementary Information).
For SN~2018zd, we identify a faint progenitor candidate that may be consistent with an SAGB star (Extended Data Figs.~\ref{EDfig:discovery} and \ref{EDfig:progenitor_sed}), and the explosion energy is consistent within the light-curve degeneracy (Extended Data Fig.~\ref{EDfig:LCscaling}).
\label{EDfig:ECSNcan}
}
\end{EDfigure}

\newpage

\begin{EDfigure}
\centering
\includegraphics[width=0.68\textwidth]{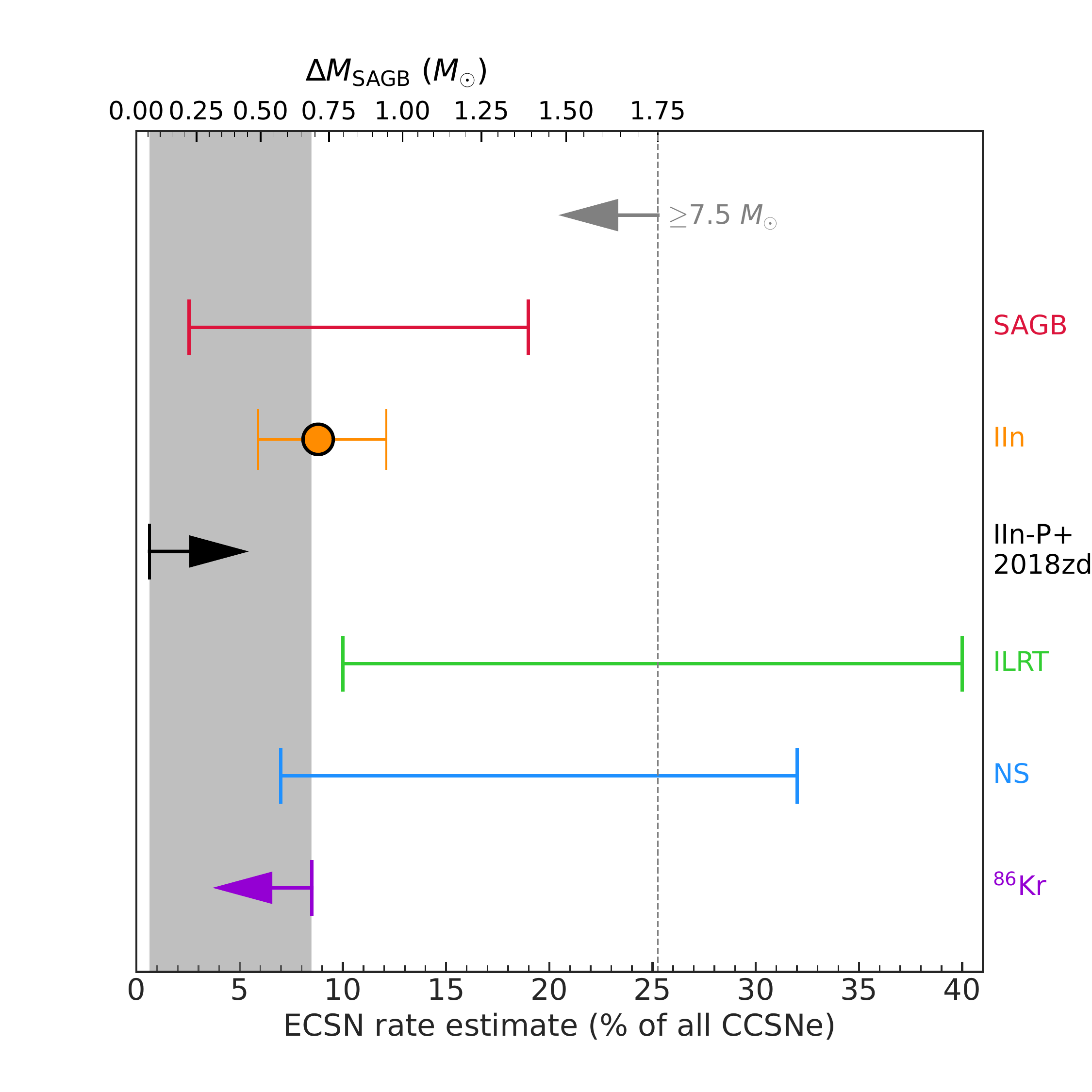}
\caption{
\textbf{ECSN rate estimators.}
Comparison of the ECSN rate estimates: 
`SAGB' is the SAGB mass window from stellar evolutionary calculations at solar metallicity\cite{Poelarends2008}; 
`IIn' is the observed Type~IIn SN rate from a volume-limited ($\leq60$\,Mpc) sample\cite{Smith2011}; 
`IIn-P+2018zd' is a rough lower limit of the Type~IIn-P SN rate within $60$\,Mpc combined with SN~2018zd (Methods); 
`ILRT' is a rough estimate from ILRTs within $30$\,Mpc\cite{Thompson2009}; 
`NS' is an estimated rate from the bimodality in the neutron star mass distribution\cite{Schwab2010} assuming that the low-mass and high-mass peaks originate from ECSNe and Fe CCSNe, respectively; 
and `$^{86}$Kr' is an upper limit from the ECSN nucleosynthesis calculation\cite{Wanajo2018} assuming that ECSNe are the dominant production source of $^{86}$Kr. 
The conversion between the fraction of all CCSNe and the SAGB mass window is performed assuming the Salpeter initial mass function with lower and upper CCSN mass limits of $7.5\,M_\odot$ and $120\,M_\odot$ (respectively) and maximum and minimum SAGB masses of $9.25\,M_\odot$ and $9.25\,M_\odot - \Delta M_{\rm{SAGB}}$ (respectively) at solar metallicity\cite{Poelarends2008}. 
The grey vertical dotted line is where the minimum SAGB mass equals the assumed lower CCSN mass limit of $7.5\,M_\odot$. 
The grey shaded region shows a rough ECSN rate constraint by the IIn-P+2018zd lower limit and the nucleosynthesis upper limit.
\label{EDfig:rate}
}
\end{EDfigure}

\end{addendum}


\clearpage

\begin{si}

\subsection{Follow-up imaging.}
\label{sec:imaging}

Follow-up imaging was obtained with the Las Cumbres Observatory network of 0.4\,m, 1\,m, and 2\,m telescopes\cite{Brown2013} through the Global Supernova Project, the Neil Gehrels \textit{Swift} Observatory Ultraviolet/Optical Telescope (UVOT), the Noguchi Astronomical Observatory (Chiba, Japan) 0.26\,m telescope, and the Itagaki Astronomical Observatory (Okayama and Tochigi, Japan) 0.35\,m and 0.5\,m telescopes. 
For the Las Cumbres photometry, PSF fitting was performed using \texttt{lcogtsnpipe}\cite{Valenti2016}, a PyRAF-based photometric reduction pipeline. {\it UBV}- and {\it gri}-band data were calibrated to Vega\cite{Stetson2000} and AB\cite{Albareti2017} magnitudes, respectively, using standard fields observed on the same night by the same telescope as the SN. 
The \textit{Swift} UVOT photometry was conducted using the pipeline for the \textit{Swift} Optical Ultraviolet Supernova Archive (SOUSA)\cite{Brown2014}, including the updated sensitivity corrections and zeropoints\cite{Breeveld2011} and the subtraction of the underlying host-galaxy count rates using images from October/November 2019.
The unfiltered optical Itagaki (KAF-1001E CCD) and Noguchi (ML0261E CCD) photometry was extracted using \texttt{Astrometrica}\cite{Raab2012} and calibrated to the Fourth US Naval Observatory CCD Astrograph Catalog (UCAC4)\cite{Zacharias2013}. 
All photometry will be available for download via WISeREP and the Open Supernova Catalog.
We correct all photometry for the Milky Way (MW) and host-galaxy extinction (Extended Data Fig.~\ref{EDfig:allLC}).

We estimate an explosion epoch by fitting a quadratic function $F_1(t-t_0)^2$ to the unfiltered Itagaki and first three Noguchi points with similar CCD spectral responses ($\lambda_{\rm eff}=6500$--$6700$\,\AA), where the effect of CSM interaction is less prominent than in the UV bands (Extended Data Fig.~\ref{EDfig:allLC}). This yields an explosion epoch $t_0 = \text{MJD} \, 58178.4\pm0.1$, where the uncertainty is estimated from the difference between the explosion epoch and the first Itagaki detection. 
Even if we use the most conservative explosion epoch of the last nondetection on $\text{MJD} \, 58175.5$, the difference is only $2.9$ rest-frame days, not affecting the main results of this paper. 

We fit a blackbody SED to every epoch of the Las Cumbres and \textit{Swift} photometry containing at least three filters (excluding the {\it r} band owing to strong H$\alpha$ contamination) obtained within $0.3$\,d of each other to estimate the blackbody temperature and radius at the assumed luminosity distance (note that the observed SED peaks are bluer than the {\it Swift} wavelength coverage $3$--$10$\,d after the explosion, potentially underestimating the blackbody temperatures\cite{Valenti2016}). Then we integrate the fitted blackbody SED to obtain bolometric (and pseudobolometric) luminosity at each epoch. 
Since we only have the unfiltered Noguchi photometry during the plateau drop owing to the Sun constraint, we estimate a bolometric (and pseudobolometric) correction by finding the offset of the Noguchi photometry to the Las Cumbres and \textit{Swift} integrated bolometric (and pseudobolometric) luminosity during the plateau phase (50--80 d) where most of the SED ($\sim80$\%) is in the spectral response range of the unfiltered CCD. Then we apply the bolometric (and pseudobolometric) correction to the Noguchi photometry and include it in the bolometric (and pseudobolometric) light curve during the plateau drop (Fig.~\ref{fig:L50}). This procedure is also justified by the good agreement with the tail bolometric (and pseudobolometric) luminosity obtained from the Las Cumbres multiband photometry after the Sun constraint.

\subsection{Follow-up spectroscopy.}
\label{sec:spec}

Follow-up spectra were obtained with the FLOYDS spectrograph mounted on the Las Cumbres Observatory 2\,m Faulkes Telescope North (FTN)\cite{Brown2013} through the Global Supernova Project, the Boller \& Chivens (B\&C) spectrograph mounted on the 2.3\,m Bok telescope, the Blue Channel (BC) spectrograph mounted on the 6.5\,m MMT, and the Low Resolution Imaging Spectrometer (LRIS)\cite{Oke1995, McCarthy1998, Rockosi2010} and the DEep Imaging Multi-Object Spectrograph (DEIMOS)\cite{DEIMOS2003} mounted on the 10\,m Keck-I and Keck-II telescopes, respectively.
For the FLOYDS observations, a $2''$-wide slit was placed on the target at the parallactic angle\cite{Filippenko1982} (to minimise the effects of atmospheric dispersion). One-dimensional spectra were extracted, reduced, and calibrated following standard procedures using \texttt{floyds\_pipeline}\cite{Valenti2014}.
The Bok low-resolution optical spectra were taken with the 300 lines mm$^{-1}$ grating using a $1.5''$-wide slit, and the MMT moderate-resolution spectra were obtained using a $1.0''$-wide slit. The spectra were reduced using standard techniques in IRAF, including bias subtraction, flat-fielding, and sky subtraction. Flux calibration was done with spectrophotometric standard star observations taken on the same night at similar airmass.
The Keck LRIS spectra were reduced using the \texttt{Lpipe} pipeline\cite{Perley2019} with the default parameters and standard spectroscopic reduction techniques. 
The Keck DEIMOS spectrum was reduced with a custom-made Python pipeline that performs flat-field correction, sky subtraction, optimal extraction\cite{Horne1986}, and flux calibration using a standard star observed on the same night as the SN.
All spectra will be available for download via WISeREP and the Open Supernova Catalog.
We correct all spectra for the MW and host-galaxy extinction and calibrate the flux using the photometry (Extended Data Fig.~\ref{EDfig:allspec}).

We measure expansion velocities of H$\alpha$, H$\beta$, and Fe~{\sc ii} $\lambda$5169 from the absorption minimum by fitting a P~Cygni profile to each line in the spectra (Extended Data Fig.~\ref{EDfig:allspec}). Then we translate the difference between the observed minimum and the rest wavelength of the line to an expansion velocity using the relativistic Doppler formula (Extended Data Fig.~\ref{EDfig:allvel}). We estimate the velocity uncertainties by randomly varying the background region by $\pm5\,\Angstrom$. 

We simultaneously fit Gaussian functions to He~{\sc i} $\lambda$7065, [Fe~{\sc ii}] $\lambda$7155, [Fe~{\sc ii}] $\lambda$7172, [Ca~{\sc ii}] $\lambda$7291, [Ca~{\sc ii}] $\lambda$7323, [Ni~{\sc ii}] $\lambda$7378, [Fe~{\sc ii}] $\lambda$7388, [Ni~{\sc ii}] $\lambda$7412, and [Fe~{\sc ii}] $\lambda$7452 in the nebular spectra assuming a single full width at half-maximum intensity (FWHM) velocity for all lines and the theoretically expected line ratios for the [Ca~{\sc ii}], [Fe~{\sc ii}], and [Ni~{\sc ii}] lines\cite{Jerkstrand2015} (Fig.~\ref{fig:nebular}). The resultant [Ni~{\sc ii}] $\lambda$7378/[Fe~{\sc ii}] $\lambda$7155 intensity ratios and FWHM velocities are $1.3$--$1.6$ and 2,500--2,100\,km\,s$^{-1}$, respectively, from 278 to 600\,d after the explosion.

\subsection{Follow-up spectropolarimetry.}
\label{sec:specpol}

Follow-up spectropolarimetric observations of SN~2018zd were obtained using the CCD Imaging/Spectropolarimeter (SPOL\cite{Schmidt1992}) on the 6.5\,m MMT telescope using a $2.8''$ slit on 2018 April 23 (53\,d after the explosion). We used a 964 lines mm$^{-1}$ grating with a typical wavelength coverage of 4,050--7,200\,{\AA} and a resolution of $\sim 29$\,{\AA}. We used a rotatable semi-achromatic half-wave plate to modulate incident polarization and a Wollaston prism in the collimated beam to separate the orthogonally polarized spectra onto a thinned, anti-reflection-coated $800\times1200$ pixel SITe CCD.  The efficiency of the wave plate as a function of wavelength was measured and corrected for by inserting a fully-polarizing Nicol prism into the beam above the slit. A series of four separate exposures that sample 16 orientations of the wave plate yield two independent, background-subtracted measures of each of the linear Stokes parameters, $Q$ and $U$. Two such sequences were acquired and combined to increase the signal-to-noise ratio.

Our spectropolarimetric analysis is performed primarily using the normalised linear Stokes parameters, $q=Q/I$ and $u=U/I$, which are rotated with respect to each other, allowing us to decompose the polarization signal into orthogonal components in position-angle space.  We use the debiased polarization level, $p_{\rm db} = \sqrt{|(q^2 + u^2) - \frac{1}{2}(\sigma_q^2 + \sigma_u^2)|}$, in favour of the standard polarization level, $p = \sqrt{q^2 + u^2}$, because the standard polarization level is a positive-definite quantity that measures the distance from the origin in a $q$ vs. $u$ plane. When the signal-to-noise ratio is low, this positive-definite quantity can be misleading, whereas the debiased polarization value accounts for large uncertainty in measurements of $q$ and $u$.

SN~2018zd exhibits a mean polarization of $0.9$\% across the continuum at 5,100--5,700\,{\AA} and $0.8$\% across the continuum at 6,000--6,300\,{\AA}.  However, the polarization does not vary much across the entire spectrum, even across absorption and emission-line features. Typically, a polarized continuum would become depolarized across emission-line features owing to dilution with unpolarized light from the emission line. Since SN~2018zd does not exhibit any such changes across any of its emission-line features, we suggest that the majority of the polarization signal arises in the interstellar medium rather than in the SN itself. The Serkowski relation\cite{Serkowski1975} suggests that $p_{\rm max} < 9\,E(B-V)$.  If all 0.9\% of the continuum peak polarization in SN~2018zd were due to the interstellar medium, then we could estimate the extinction to be $E(B-V) > p_{\rm max}/9 = 0.1$\,mag and a reddening of at least $A_V = 3.1\,E(B-V) = 0.31$\,mag.

\subsection{Extra \texttt{MESA}+\texttt{STELLA} modeling description.}
\label{sec:MESAdet}

All progenitor models began at solar metallicity ($Z=0.02$), and the naming scheme gives progenitor and explosion properties as follows: ($\mathrm{M}[M_\mathrm{ej}/M_\odot]\_\mathrm{R}[R/R_\odot]\_\mathrm{E}[E_\mathrm{exp}/10^{51}\,\mathrm{erg}]$). 
The high-ejecta-mass model, M17.2\_R718\_E0.48, is $18.8\,M_\odot$ at core collapse ($20\,M_\odot$ at zero-age main sequence (ZAMS)) with no rotation, no exponential overshooting ($f_\mathrm{ov}=f_{0,\mathrm{ov}}=0.0$), mixing length $\alpha_\mathrm{env}=2.0$ in the H-rich envelope, and a wind efficiency factor $\eta_\mathrm{wind}=0.4$.
The moderate model, M14.5\_R864\_E0.37, is $16.3\,M_\odot$ at core collapse ($17\,M_\odot$ at ZAMS) with modest initial rotation $\Omega/\Omega_\mathrm{crit} = 0.2$, no exponential overshooting, $\alpha_\mathrm{env}=2.0$, and $\eta_\mathrm{wind}=0.2$. 
The low-ejecta-mass and large-radius model, M8.3\_R1035\_E0.23, is $11.8\,M_\odot$ at core collapse ($15\,M_\odot$ at ZAMS) with modest rotation $\Omega/\Omega_\mathrm{crit} = 0.2$, moderately high exponential overshooting ($f_\mathrm{ov}=0.018, f_{0,\mathrm{ov}}=0.006$), $\alpha_\mathrm{env}=2.0$, and $\eta_\mathrm{wind}=0.9$. Despite the ZAMS mass typical of an RSG, this model sufficiently captures the relevant explosion properties for the SAGB explosion scenario, as the mass of the H-rich ejecta, explosion energy, and progenitor radius determine the plateau properties of Type~II-P SNe, not the ZAMS mass.

In \texttt{MESA} revision 12115, a thermal bomb was injected in the innermost $0.1\,M_\odot$ of each model, heating the star to the desired total final energy $E_\mathrm{exp}$, with the updated prescription for removing material falling back onto the inner boundary\cite{Paxton2019, Goldberg2019}, which can be relevant at the low explosion energies required here.
Of the three explosions, only M8.3\_R1035\_E0.23 undergoes substantial late-time fallback, totaling $2\,M_\odot$, which is excised from the model with no extra heating and negligible change in the total explosion energy. The evolution of the shock was modelled in \texttt{MESA} with the `Duffell RTI' prescription for mixing via the Rayleigh-Taylor instability\cite{Duffell2016,Paxton2018}, terminating near shock breakout, when the shock reaches a mass coordinate of $0.04\,M_\odot$ below the outer boundary of each model. The $^{56}$Ni distribution in each model was then scaled to match the observed value of $0.0086\,M_\odot$. 
Then in \texttt{STELLA}, bolometric light curves and expansion velocities were produced using 600 spatial zones and 100 frequency bins, without any additional material outside the stellar photosphere. For models with CSM, 600 zones are used in \texttt{STELLA}, including 400 zones for the original ejecta, and 200 additional zones for the wind model.

\subsection{Host galaxy.}
\label{sec:NGC2146}

NGC~2146 is an edge-on spiral galaxy with several tidal streams that were probably ejected during a galaxy merger event $\sim 800$\,Myr ago\cite{Taramopoulos2001} (Extended Data Fig.~\ref{EDfig:discovery}). 
The presence of a starburst-driven superwind from the bulge is revealed across the electromagnetic spectrum from $\gamma$-rays to infrared\cite{Tang2014, Armus1995, Taramopoulos2001, Kreckel2014}, indicating an ongoing high star-formation rate\cite{Skibba2011} (SFR $\approx 10\,M_\odot$\,yr$^{-1}$). 
On the basis of radio observations of the bulge\cite{Tarchi2000}, there are many more dense H~{\sc ii} regions (each containing up to 1000 type O6 stars) than supernova remnants, suggesting a relatively young phase of the starburst. 
The bulge has a high dust content and roughly solar metallicity ($12+ {\rm log}_{10}[{\rm O/H}] = 8.68\pm0.10$)\cite{Skibba2011,Aniano2020}. 
Since SN~2018zd is at a relatively large separation from the nucleus of $1{\farcm}83$ northwest ($36{\farcs}1$ north, $103{\farcs}7$ west; Extended Data Fig.~\ref{EDfig:discovery}), and the galactic radius parameter $R_{25}=1{\farcm}78$ (via the NASA/IPAC Extragalactic Database), if we reasonably assume that there is an abundance gradient for the galaxy, the metallicity at the SN site is probably subsolar; this merits future investigations once the SN fades.

\subsection{Alternative scenarios.}
\label{sec:alternatives}

A low-mass ($\lesssim9.6\,M_\odot$) Fe CCSN is a possible alternative for SN~2018zd, as similar explosion energy ($\sim10^{50}$\,erg)\cite{Muller2019} and nucleosynthesis\cite{Wanajo2018} to ECSNe may be expected because of a similar steep density gradient outside the degenerate core. 
For a low-mass RSG star, however, no high constant ($\gtrsim10^{-5}\,M_\odot$\,yr$^{-1}$)\cite{Mauron2011, Goldman2017, Beasor2020} and/or eruptive\cite{Smartt2009, Fuller2017} mass loss is expected to produce dense confined He-, C-, and N-rich, but O-poor CSM (but note that the mass loss is quite sensitive to the model treatments of, for example, convection and off-center nuclear burning\cite{Woosley2015}). 
In addition, a low-mass RSG has Si-, O-, and He-rich layers\cite{Jones2013} which are expected to produce additional Si, S, Ca, Mg, O, C, and He lines in nebular spectra\cite{Jerkstrand2018}.
Thus, a low-mass Fe CCSN may be able to explain the light-curve morphology, but probably not the early-time CSM interaction and nebular spectra observed for SN~2018zd.

On the other side of the progenitor mass spectrum, another possible alternative for SN~2018zd is a high-mass ($\gtrsim25\,M_\odot$) Fe CCSN, as small kinetic energy ($\sim10^{50}$\,erg) and ejected radioactive $^{56}$Ni mass ($\lesssim10^{-3}\,M_\odot$) may be expected owing to fallback accretion onto the central remnant\cite{Lisakov2018,Chan2018}. 
For such high fallback accretion, however, extra luminosity ($L \propto t^{-5/3}$) at late times ($t\gtrsim200$\,d) is expected\cite{Dexter2013,Moriya2019}. 
Also, no ejected stable $^{58}$Ni should be observed, as it is produced in the innermost neutron-rich layer\cite{Wanajo2009}. 
Thus, a high-mass Fe CCSN may be able to explain the photospheric light curve, but not the late-time exponential tail and nebular spectra of SN~2018zd.

If the luminosity distance to NGC~2146 were larger than $12$\,Mpc, it would be quite unlikely that SN~2018zd is an ECSN, since $M_{\rm Ni} > 0.01\,M_\odot$, $E_{\rm exp} > 4\times10^{50}$\,erg, and $M_{\rm ej} > 10\,M_\odot$ in a reasonable progenitor radius range of $400$--$1400\,R_\odot$ according to the light-curve scaling (Eq.~\ref{eq:EandMofR} in Methods). Then SN~2018zd would become a real challenge to stellar evolution and SN explosion theories to reconcile all of the observational ECSN indicators with a higher $M_{\rm Ni}$, $E_{\rm exp}$, $M_{\rm ej}$, and $M_{\rm ZAMS}$ for the progenitor.
If the luminosity distance were $18$\,Mpc, the progenitor candidate detection of SN~2018zd in \textit{HST} F814W would become as bright as that of the SN~2005cs progenitor (Extended Data Fig.~\ref{EDfig:progenitor_sed}), but still on the faint end of Type~II SN progenitors\cite{Smartt2009, Smartt2015} despite the expected higher $M_{\rm ej}$ and $M_{\rm ZAMS}$ from the light-curve scaling.

We note that Zhang et al. (2020)\cite{Zhang2020} also discusses a possible ECSN origin for SN~2018zd based on the small radioactive $^{56}$Ni yield, dense CSM, and faint X-ray radiation. Owing to their adopted larger luminosity distance ($18.4\pm4.5$\,Mpc; Methods), however, they suggest that SN~2018zd is a member of the class of luminous Type~II SNe with low expansion velocities\cite{Rodriguez2020}, which probably arise from extended CSM interaction ($4$--$11$ weeks after the explosion).
In this work, we perform numerical light-curve modeling and demonstrate that ECSN-parameter explosions with the early CSM interaction ($\sim30$\,days after the explosion) can reproduce both the light-curve and velocity evolution (Fig.~\ref{fig:uvUV} and Extended Data Fig.~\ref{EDfig:LCscaling}). Furthermore, we present the progenitor candidate identification (Extended Data Figs.~\ref{EDfig:discovery} and \ref{EDfig:progenitor_sed}) and more detailed spectral analyses (Figs.~\ref{fig:flash} and \ref{fig:nebular}), showing that the chemical composition and nucleosynthesis are consistent with those expected for ECSNe.

\subsection{Other ECSN candidates.}
\label{sec:candidates}

SN~1054, whose remnant is the Crab Nebula, has been suggested as an ECSN candidate\cite{Nomoto1982, Nomoto1984, Nomoto1987, Smith2013, Tominaga2013, Moriya2014}. 
It shows He-, C-, and Ni-rich ejecta, but O- and Fe-poor abundances\cite{Hudgins1990,Satterfield2012}, small ejecta mass ($4.6\pm1.8\,M_\odot$)\cite{Fesen1997}, and low kinetic energy ($\sim 10^{49}$\,erg)\cite{Smith2013}. The slowly expanding filaments ($\sim1200$\,km\,s$^{-1}$) without a blast wave outside probably indicate the presence of CSM decelerating the SN ejecta\cite{Fesen1997, Smith2013}, and the historical light curve of SN~1054 may be similar to that of ECSNe\cite{Smith2013,Tominaga2013,Moriya2014}. 
However, the observed relatively high neutron star kick velocity ($\sim 160$\,km\,s$^{-1}$) is at odds with those theoretically predicted for ECSNe ($<10$\,km\,s$^{-1}$)\cite{Gessner2018}.
On the other hand, the pre-collapse O+Ne+Mg core of an SAGB star could have large rotation and even `super-Chandrasekhar' mass if the angular momentum transport from the rotating core to the very extended SAGB envelope is small during contraction\cite{Uenishi2003,Benvenuto2015,Hachisu2012}. The collapse of such an unstable core could in principle yield a large spin and kick. 

In addition to SN~1054, other previously suggested ECSN candidates can be divided into three main types: intermediate-luminosity red transients\cite{Prieto2008, Botticella2009, Thompson2009, Adams2016, Cai2018, Stritzinger2020} (ILRTs; for example, SN~2008S and AT~2017be), low-luminosity Type~II-P SNe\cite{Kitaura2006, Janka2008, Spiro2014, Hosseinzadeh2018}, and Type~IIn-P SNe (for example, SNe~2009kn and 2011ht)\cite{Kankare2012, Mauerhan2013, Moriya2014, Smith2013} (Extended Data Fig.~\ref{EDfig:ECSNcan}).

ILRTs are the luminosity gap transients between novae and SNe, whose origin has been debated as either a massive-star outburst\cite{Bond2009, Berger2009, Smith2009, Smith2011b} or a terminal ECSN explosion\cite{Prieto2008, Botticella2009, Thompson2009, Adams2016, Cai2018, Stritzinger2020}. Their progenitors are surrounded by dusty, optically thick shells, resulting in CSM-dominated transients\cite{Prieto2008, Botticella2009, Thompson2009, Bond2009, Berger2009, Adams2016}. However, their faint light-curve morphology with CSM interaction requires extremely low explosion energy ($\lesssim10^{48}$ erg) that is unexpected for ECSNe\cite{Tominaga2013, Moriya2014, Moriya2016}, and their chemical composition and nucleosynthesis are unclear owing to the lack of nebular-phase spectra.

Low-luminosity Type~II-P SNe typically yield low $^{56}$Ni mass ($\lesssim 10^{-2}\,M_\odot$)\cite{Spiro2014} with ECSN-like light-curve morphology (Fig.~\ref{fig:L50}). However, their chemical composition and nucleosynthesis are inconsistent with ECSNe\cite{Jerkstrand2018} (Supplementary Fig.~\ref{SIfig:LL_neb}), and their CSM density is generally low compared with that expected from ECSNe\cite{Poelarends2008, Moriya2014} (except for SN~2016bkv\cite{Hosseinzadeh2018}). Low-mass RSG progenitors have been directly identified for SNe~2003gd\cite{Maund2014a}, 2005cs\cite{Eldridge2007, Maund2014a}, and 2008bk\cite{Maund2014b}, excluding SAGB stars -- the progenitors of ECSNe.

Type~IIn-P SNe show Type~IIn-like narrow CSM emission lines in spectra and Type~II-P-like light-curve morphology with large plateau drops similar to ECSNe\cite{Kankare2012, Mauerhan2013, Smith2013, Moriya2014} (Supplementary Fig.~\ref{SIfig:IIn-P}). 
The SN signatures (for example, chemical abundance) are mostly hidden below the CSM interaction, in general. For Type~IIn-P SN~2011ht\cite{Humphreys2012}, however, we measure [Ni~{\sc ii}] $\lambda$7378/[Fe~{\sc ii}] $\lambda$7155 $= 3.8$ at $155$\,d after the explosion (using a public spectrum on WISeREP), which may indicate ECSN-like nucleosynthesis, although the spectrum may not be fully nebular given the relatively early phase. While no SN~IIn-P progenitors have been directly identified, a pre-explosion outburst has been observed for SN~2011ht\cite{Fraser2013}. The true nature of Type~IIn-P SNe is yet to be revealed.

%


\subsection{Supplementary References}


\clearpage

\subsection{Supplementary Figures}

\begin{SIfigure}
\centering
\includegraphics[width=0.99\textwidth]{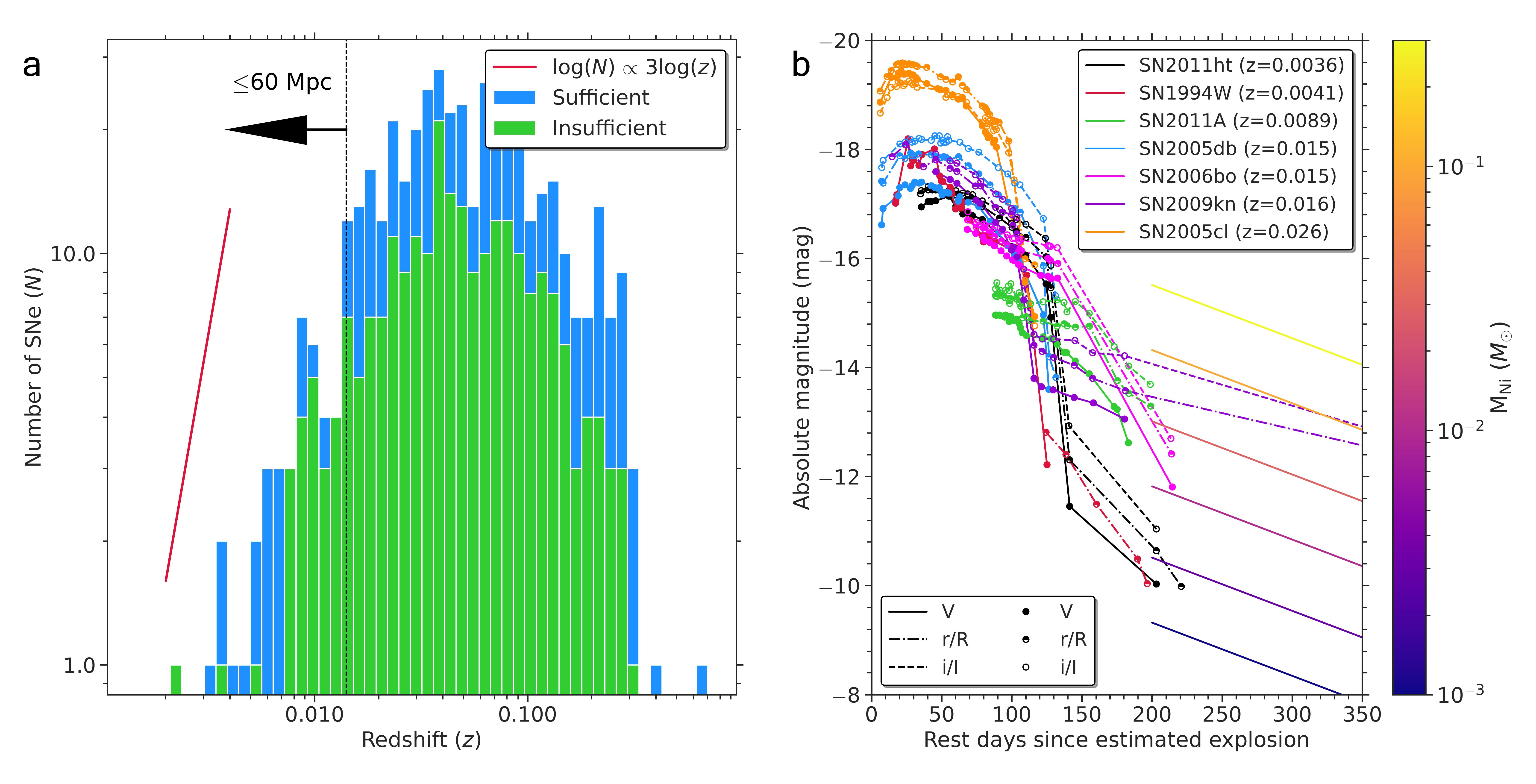}
\caption{
\textbf{Public Type~IIn and IIn-P SN samples.}
\textbf{a,} Redshift distribution of the 455 public Type~IIn SNe retrieved from WISeREP and/or TNS. 
241 objects have insufficient public spectra and/or light curves to secure the Type~IIn classifications and/or to be identified as Type~IIn-P SNe, but are included in the sample so as not to overestimate the lower limit. 
The red line is the number-density slope by assuming the volume term with the standard cosmology (H$_0=71.0$ km\,s$^{-1}$\,Mpc$^{-1}$, $\Omega_{\Lambda_0}=0.7$, and $\Omega_{m_0}=0.3$, giving $d_L \propto z$ for $z < 0.1$). 
The black dotted line is the distance cut ($\leq60$\,Mpc) we apply to compare with the LOSS sample\cite{Smith2011}. 
By comparing the number-density slope to the sample histogram as a first-order estimation, the sample does not seem to suffer substantially from incompleteness within $60$\,Mpc. 
\textbf{b,} Comparison of the identified Type~IIn-P SN candidates by applying the two light-curve criteria. 
The explosion epochs of SNe~2006bo and 2011A are not well constrained and can shift up to $\pm64$\,d and $\pm85$\,d, respectively\cite{Taddia2013,deJaeger2015}. 
The colour-coded tails at 200--350\,d are the expected $V$-band tails from the fully trapped radioactive heating for a given $^{56}$Ni mass\cite{Hamuy2003}. 
The observed Ni-mass upper limits are within $10^{-3}$ to $3\times10^{-2}\,M_\odot$, assuming that the tails are purely powered by the radioactive heating.
\label{SIfig:IIn-P}
}
\end{SIfigure}

\newpage

\begin{SIfigure}
\centering
\includegraphics[width=0.99\textwidth]{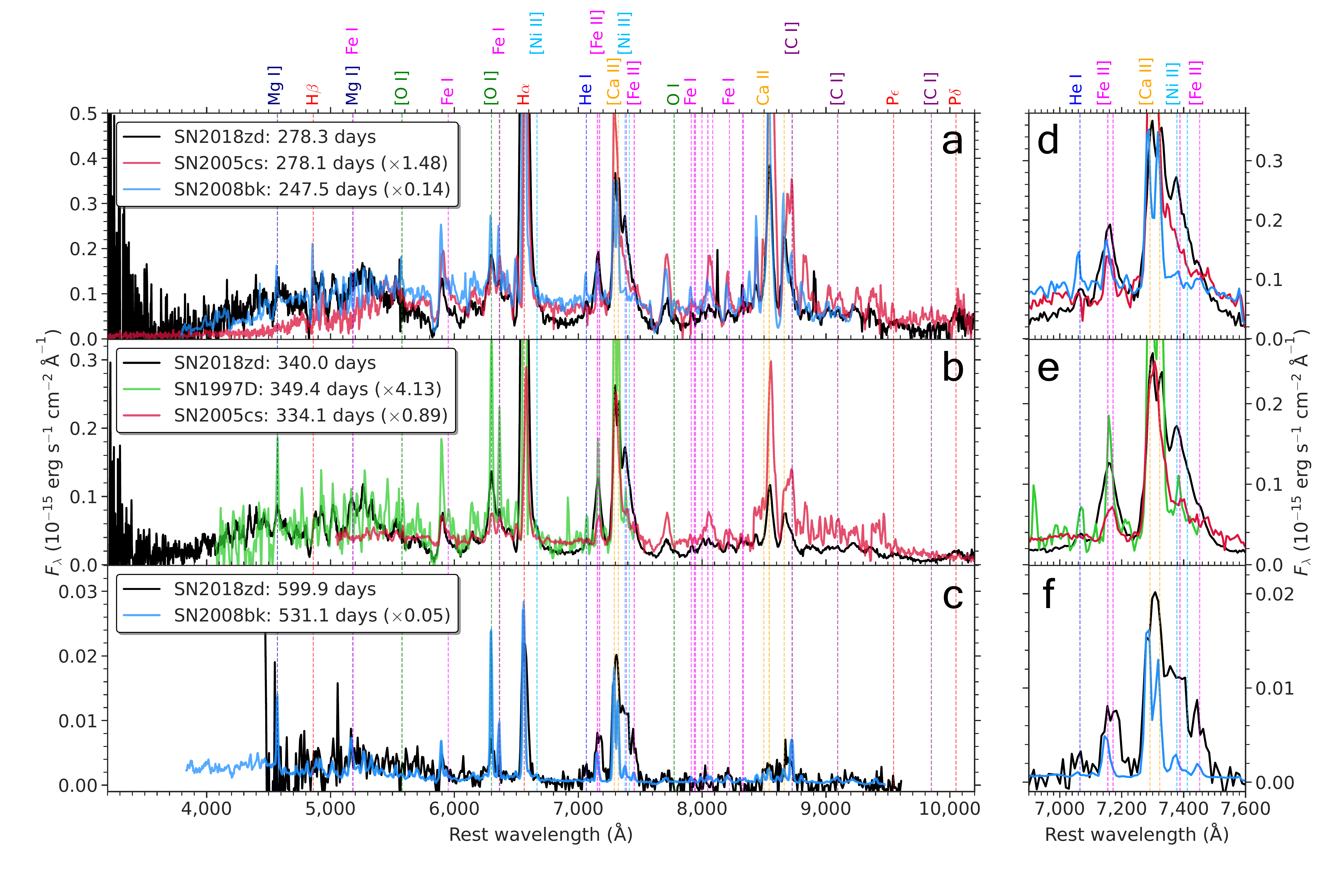}
\caption{
\textbf{Nebular spectral time series of low-luminosity Type~II-P SNe.}
\textbf{a--c,} Comparison of the nebular spectral time series of SN~2018zd with the scaled (by integrated flux as in the legend) and resampled low-luminosity Type~II-P SNe 1997D\cite{Benetti2001}, 2005cs\cite{Pastorello2009}, and 2008bk\cite{VanDyk2012,Maguire2012,Gutierrez2017}.
In ascending order of wavelength, note the distinct Mg~{\sc i}] $\lambda$4571 and [O~{\sc i}] $\lambda\lambda$6300, 6364 + Fe~{\sc i} $\lambda$6364 observed in SN~1997D; 
Fe~{\sc i} cluster 7,900--8,500\,\AA, [C~{\sc i}] $\lambda$8727, and [C~{\sc i}] $\lambda$9100 observed in SN~2005cs; 
and Mg~{\sc i}] $\lambda$4571, [O~{\sc i}] $\lambda\lambda$6300, 6364 + Fe~{\sc i} $\lambda$6364, Fe~{\sc i} cluster 7,900--8,500\,\AA, and [C~{\sc i}] $\lambda$8727 observed in SN~2008bk.
\textbf{d--f,} Same as panels (\textbf{a})--(\textbf{c}), but zoomed into the wavelength range of interest (as in Fig.~\ref{fig:nebular}). 
Note the line-intensity ratios of [Ni~{\sc ii}] $\lambda$7378/[Fe~{\sc ii}] $\lambda$7155 $< 1$ observed in SNe~1997D, 2005cs, and 2008bk.
The strong C, O, Mg, and/or Fe lines combined with the weak Ni lines observed in SNe~1997D, 2005cs, and 2008bk are inconsistent with the ECSN chemical composition and nucleosynthesis.
\label{SIfig:LL_neb}
}
\end{SIfigure}

\end{si}

\end{document}